\documentclass[%
 aip,
 amsmath,amssymb,
 reprint,%
]{revtex4-1}

\usepackage{graphicx}% Include figure files
\usepackage{dcolumn}% Align table columns on decimal point
\usepackage{bm}% bold math
\usepackage[utf8]{inputenc}
\usepackage[T1]{fontenc}
\usepackage{mathptmx}
\usepackage{etoolbox}
\usepackage{comment}
\usepackage{caption}
\usepackage{subcaption}

\makeatletter
\def\@email#1#2{%
 \endgroup
 \patchcmd{\titleblock@produce}
  {\frontmatter@RRAPformat}
  {\frontmatter@RRAPformat{\produce@RRAP{*#1\href{mailto:#2}{#2}}}\frontmatter@RRAPformat}
  {}{}
}%
\makeatother
\begin{document}

\preprint{AIP/123-QED}

%\title[Sample title]{Sample Title:\\with Forced Linebreak}
\title{MolDStruct: modelling the dynamics and structure of matter exposed to ultrafast
X-ray lasers with hybrid
collisional-radiative/molecular dynamics}

%Large-scale hybrid molecular dynamics and photon-matter method to explore ultrafast interactions with X-ray lasers}

% Force line breaks with \\
\author{Ibrahim Dawod}
% \altaffiliation[Also at ]{Physics Department, XYZ University.}%Lines break automatically or can be forced with \\\
\affiliation{ 
Department of Physics and Astronomy, Uppsala University, Box 516, SE-751 20 Uppsala, Sweden%\\This line break forced with \textbackslash\textbackslash
}%
\affiliation{ 
European XFEL, Holzkoppel 4, DE-22869 Schenefeld, Germany
}%
\author{Sebastian Cardoch}%
\affiliation{ 
Department of Physics and Astronomy, Uppsala University, Box 516, SE-751 20 Uppsala, Sweden%\\This line break forced with \textbackslash\textbackslash
}%

\author{Tomas Andr\'e}
\affiliation{ 
Department of Physics and Astronomy, Uppsala University, Box 516, SE-751 20 Uppsala, Sweden%\\This line break forced with \textbackslash\textbackslash
}%
\author{Emiliano De Santis}
\affiliation{Department of Chemistry -- BMC, Uppsala University, Box 576, SE-751 23 Uppsala, Sweden}
\author{Juncheng E}
\affiliation{ 
European XFEL, Holzkoppel 4, DE-22869 Schenefeld, Germany
}%
\author{Adrian P. Mancuso}
\affiliation{ 
European XFEL, Holzkoppel 4, DE-22869 Schenefeld, Germany
}%
\affiliation{ 
Department of Chemistry and Physics, La Trobe Institute for Molecular Science,
La Trobe University, Melbourne, Victoria 3086, Australia
}%
\affiliation{ 
Diamond Light Source, Harwell Science and Innovation Campus, Didcot OX11 0DE, UK
}%
\author{Carl Caleman}
\affiliation{ 
Department of Physics and Astronomy, Uppsala University, Box 516, SE-751 20 Uppsala, Sweden%\\This line break forced with \textbackslash\textbackslash
}%
\affiliation{ 
Center for Free-Electron Laser Science, Deutsches Elektronen-Synchrotron, Notkestraße 85 DE-22607 Hamburg, Germany
}%

\author{Nicusor Timneanu}
\email{ibrahim.dawod@physics.uu.se, carl.caleman@physics.uu.se, nicusor.timneanu@physics.uu.se}
\affiliation{ 
Department of Physics and Astronomy, Uppsala University, Box 516, SE-751 20 Uppsala, Sweden%\\This line break forced with \textbackslash\textbackslash
}%

\date{\today}% It is always \today, today,

\begin{abstract}
We describe a method to compute photon-matter interaction and atomic dynamics with X-ray lasers using a hybrid code based on classical molecular dynamics and collisional-radiative calculations. The forces between the atoms are dynamically computed based on changes to their electronic occupations and the free electron cloud created due to the irradiation of photons in the X-ray spectrum. %The forces between the atoms in the sample are formulated by spherically symmetric two-body potentials for the bonded and non-bonded interactions. Screened potentials for the non-bonded interactions are developed based on the collisional-radiative computations. 
The rapid transition from neutral solid  matter to dense plasma phase allows the use of screened potentials, which reduces the number of non-bonded interactions required to compute. In combination with parallelization through domain decomposition, large-scale molecular dynamics and ionization induced by X-ray lasers can be followed. 
This method is applicable for large enough samples (solids, liquids, proteins, viruses, atomic clusters and crystals) that when exposed to an X-ray laser pulse turn into a plasma in the first few femtoseconds of the interaction. We show several examples of the applicability of the method and we quantify the sizes that the method is suitable for. 
For large systems, we investigate non-thermal heating and scattering of bulk water, which we compare to previous experiments. We simulate molecular dynamics of a protein crystal induced by an X-ray pump, X-ray probe scheme, and find good agreement of the damage dynamics with experiments.  For single particle imaging, we simulate ultrafast dynamics of a methane cluster exposed to a femtosecond X-ray laser. In the context of coherent diffractive imaging we study the fragmentation as given by an X-ray pump X-ray probe setup to understand the evolution of radiation damage.% and the effect of electron-ion coupling. 

\end{abstract}

\maketitle

\section{\label{sec:level1}INTRODUCTION\protect\\}
X-rays are routinely used to study biological systems like proteins, viruses or cells. X-ray established methods provide a three-dimensional view of their structures, aiding in a deeper understanding of how they function. Synchrotrons have for many decades allowed the structure determination of biomolecules, but it requires them to be in a crystallized form \cite{ProteinCrystallography}. For non-crystalline samples, cryogenic electron microscopy emerged as the main imaging tool\cite{cheng2015primer}. These methods have achieved atomic resolution structures but are  unable to study time-resolved dynamics of biology on the femtosecond time-scale \cite{frank2017time}, or even on the picosecond timescale. The X-ray free-electron laser (XFEL) provides pulses with high intensity on femtosecond timescales, \cite{mancuso2019single} and have recently been able to follow the femtosecond resolved dynamics at atomic resolution\cite{nass2020structural, christou2023time}. %These lasers provide pulses with high intensity on femtosecond timescales \cite{mancuso2019single}. %The large energy deposited in the sample leads to structural changes, while structural information has been obtained through coherent diffraction . 
The appreciable energy deposited by the X-ray pulses
in the sample leads to measurable structural changes. Depending on the parameters of the X-ray laser and the sample, structural information can be obtained through coherent diffraction with minimal structural alteration \cite{neutze2000potential}. 
%but using a short-femtosecond pulse near-native structural information can be been obtained through coher-ent diffractio
%It has however been shown, that electronic and structural changes occur on timescales shorter than commonly used pulses.
Atomic displacement during the pulse is overcome by using shorter pulses than typical time-scales for atomic movement \cite{neutze2000potential}. Electronic processes due to free electrons such as collisional ionization can also be minimized by using shorter pulses. One therefore needs a reliable tool to quantify radiation damage given a set of pulse parameters \cite{yoon2016comprehensive}. Recent studies have shown that long pulses can be used due to the self-gating effect where the signal in the later part of the pulse does not contribute to structural information \cite{barty2012self, Caleman:15, Martin:it5006}.    
%Using theory to model the photon-matter interaction has been previously done, Cimerron project etc.
Both single particle imaging \cite{martin2015single} (SPI), where a single molecule is imaged, and serial femtosecond crystallography\cite{chapman2011femtosecond} (SFX), where crystals are probed, are affected by radiation damage. For SPI, there is only a single copy of the sample in the interaction region and thus the studied diffraction pattern will be sensitive to local changes in the molecule. For SFX the effect of damage is less severe since there are a great number of copies of the sample which makes it so that non-reproducible motion induced by the photons will not contribute to the Bragg spots. However, if one wishes to achieve resolutions below 1 \AA{}, which is required to follow detailed electron density changes \cite{hirano2016charge} in systems like metalloproteins, damage needs to be quantified. Even in the case of negligible atomic displacement due to the photon-matter interaction, it is important to include non-neutral charge states in the reconstruction process as it could provide a molecular model that fits the data better.
%which fits the structure that is imaged better. 
%ignoring non-neutral charge states in the reconstruction process might provide a structure further from the native sample. 

Theoretical studies focused on modelling the effects of radiation damage are important to asses the limits of an experiment and can provide a guide for which parameters to use in order to gain optimal results \cite{yoon2016comprehensive}. As SPI and SFX methods are continuously developed, there are several emerging methods at XFELs for structural determination such as fluctuation X-ray scattering \cite{FXS} and X-ray fluorescence intensity correlations \cite{FIC}, where the ultrafast dynamics and ionization during the pulse plays a significant role. At the extreme in imaging experiments, the sample reaches the level of warm dense matter\cite{HEDScience}, a strongly coupled system at solid density which is quite unique on Earth, and the exploration of such transient states is now accessible at XFELs, as well as validation of the theoretical models used to describe it.

Several different published models are available for calculating radiation doses in protein crystallography\cite{RADDOSE-XFEL} and modelling photon-matter interaction and atomic dynamics for XFELs\cite{cimarron, jurek2016xmdyn, ho2017large, cryst10060478}. These include models that explicitly follow both the ion and electron dynamics with time \cite{jurek2016xmdyn, ho2017large}. However, this requires very short time-steps to resolve the fast movement of the electrons, which significantly slows down the simulations, and puts computational limits on the size of the systems that can be simulated. On the contrary, models where electrons are instead treated as a continuum have been developed \cite{cryst10060478}, which reduces the computational time of the simulation. 
We describe below a hybrid method with a similar approach, and furthermore 
%In contrast to the codes that have been published, 
our developed version of the code is openly available to be used by the community\footnote{\textsc{MolDStruct} builds on  \textsc{Gromacs}  and is available at https://github.com/moldstruct/cr-md}.

\section{METHODS}
We have developed a method to study photon-matter interaction based on a hybrid classical molecular dynamics (MD) and collisional-radiative (CR) computations, which we called \textsc{MolDStruct}. In this work, time-evolved charge distributions, ion/electron temperatures, electronic states and free electron densities are calculated using the CR code \textsc{Cretin}\cite{scott1994a}. However, any other code that outputs this information can be used. For propagating the positions of the atoms, we use a modified version of the classical MD program \textsc{Gromacs} version 4.5\cite{10.1093/bioinformatics/btt055}. 
With this version, vacuum and periodic boundary conditions can run on multiple CPU cores through domain decomposition\cite{hess}.
The strength of the CR calculations is that we can compute photon-matter interaction for arbitrarily large samples. This is because the code models the system as a continuum and it is therefore not computationally limited regarding number of particles in the system. By coupling the CR data to the MD code, we can make use of the strength of MD which gives direct information regarding the atom's position. This provides the possibility to probe local dynamics in the structure, which can be especially important for active or metallic sites in biomolecules.

\subsection{Modelling photon-matter interaction based on collisional-radiative theory}
The first step in our approach builds on a code that can provide the required photon-matter data to be coupled to the molecular dynamics simulations, and in this work we utilize the code  \textsc{Cretin}. It uses a rate-equation model based on atomic data to evolve the conditions of the material under non-local thermodynamic equilibrium environment. The atomic data includes energy levels as well as cross sections for photon and electron-induced ionization and excitations, Auger-Meitner relaxation and their inverse processes using the assumption of statistical equilibrium. At each time step, the code couples radiation transport, electronic populations, temperatures and densities in an iterative manner until reaching a self-consistent solution.  The free electrons produced are not allowed to escape. They are thermalized and described by a Maxwell-Boltzmann distribution with a single temperature.  The densities are allowed to change with hydrodynamic expansion. Environmental effects on the atom's electronic structure are included using continuum lowering, as formulated by the Stewart-Pyatt model \cite{stewart1966lowering}. 
The CR code does not include shake up/shake  off or molecular information. The model also transitions from the extension of the weakly or strongly coupled regime, which matters for the description during the first few femtoseconds following the pulse. For the simulations shown here, we utilize the screened hydrogenic model for the atoms \cite{MORE1982345}.

By utilizing an X-ray probe with specified pulse energy, duration, photon energy, and focal spot, we calculate the time-resolved interaction between photons and matter by incorporating a description of the sample, including its mass density and atomic composition (stoichiometry).
%Given an X-ray probe with a defined  pulse energy, duration, photon energy and focal spot, we can together with a description of the sample through the mass density and relative number of atoms, compute the time-resolved photon-matter interaction. 
From the CR simulation, we acquire the time-resolved fraction of each charge state for every atomic species, the free electron density and temperature, atomic number densities and electron-ion coupling. The ion and electron temperature ($T_{ion}$, $T_e$) in the CR simulations are coupled in a two-temperature model describing their evolution \cite{scott1994glf, cardoch2023decreasing}, through an electron-ion coupling $\gamma$,
\begin{equation}
    \frac{\partial T_{ion}}{\partial t}= \gamma (T_e-T_{ion}).
    \label{TTM}
\end{equation}
%The coupling between the ions and electrons will include changes in the electron temperature, which subsequently will change the rates for processes dependent electron temperature, such as electron collisional processes.
%The coupling between the ions and electrons will change their temperatures, which subsequently will change the rates for processes dependent on temperature, such as electron collisional rates.
The equilibration time between ions and electrons depends on the sample and pulse parameters used, but it can be estimated to be on the order of multiple hundreds of femtoseconds \cite{PhysRevE.83.016403}, making electron ion coupling relevant for simulations on these timescales. 
%This makes electron-ion coupling most important to include for longer simulations. %, where the ions have enough time to move. 
The effect of the expansion of the sample on the properties of the electronic distribution and the ions is included through hydrodynamic expansion \cite{scott1994a, scott1994glf}. To simulate the atoms' movement, we have adapted the \textsc{Gromacs} code\cite{10.1093/bioinformatics/btt055} such that the charge states of the atoms are set based on the CR data. We assign integer charges to individual atoms in the MD in a stochastic manner. Averaged together, these closely match the charge states form \textsc{Cretin}. We do not assign electronic states in  \textsc{Gromacs}, therefore making no distinction between valance and core ionization in the interactions.
The effect of the free electrons given by the CR calculation in the MD simulation is included through screening of the non-bonded interactions and by electron-ion coupling using Langevin temperature coupling \cite{goga2012efficient, molina2022molecular}.

The potential used to model the Coulomb interaction will depend on whether the plasma is in the weakly or strongly coupled regime. This is dependent on the relation between kinetic energy of the particles, and the Coulomb potential that they experience. 
\begin{comment}
    This is determined from the Coulomb coupling parameter $\Gamma$, which relates the average Coulomb interaction to the kinetic energy 
\begin{equation}
    \Gamma = \frac{<E_{Coulomb}>}{<E_{Kinetic}>}
\end{equation}
Given the net charge $Z_i$, elementary charge $e$, ion temperature $T_i$ and radius $R_i$ the coupling is calculated as  \cite{MURILLO19981}
\begin{equation}
    \Gamma = \frac{E_C}{k_B T_i} = \frac{<Z_ie>^2}{R_i k_B T_i}
\end{equation}
with $R_i = (\frac{3}{4\pi n_i})^{1/3}$.
\end{comment}
If the system is in the weakly coupled regime for the low charge density/high temperature case, one can assume the Coulomb potential $\phi_{C}$ is much smaller than the ion or electron temperature $T$, $\phi_{C}/k_BT << 1$, which results in the Debye-screened (D) Coulomb interaction 
\begin{equation}
    \phi_{D}(r_{ij}) = \frac{1}{4\pi \epsilon_0}\frac{q_iq_j}{r_{ij}}\exp \bigg (\frac{-r_{ij}}{\lambda_D}\bigg ),
\label{SCreend coulomb potential}
\end{equation}
where $q$ is the charge, $\epsilon_0$ the vacuum permeability and $r_{ij}$ the distance between ion $i$ and $j$. 
\begin{comment}
    
As all interactions between the atoms are formulated by electromagnetic force, screening due to free electrons will affect all interactions between two atoms. We therefore modulated both the Coulomb and LJ-interaction according to the Debye
length. For the Coulomb interaction, we get \cite{chen1984introduction}
\begin{equation}
    V_{C}(r_{ij}) = f\frac{q_iq_j}{\epsilon_0r_{ij}}\exp \bigg (\frac{-r_{ij}}{\lambda_D}\bigg ),
\label{SCreend coulomb potential}
\end{equation}
and the Lennard-Jones (LJ) interaction,
\begin{equation}
    V_{LJ}(r_{ij}) = \bigg (\frac{c_{12}}{r_{ij}^{12}} - \frac{c_6}{r_{ij}^6} \bigg ) \exp \bigg (\frac{-r_{ij}}{\lambda_D}\bigg ).
\label{Screend Lennard Jones}
\end{equation}
\end{comment}
The Debye length $\lambda_D$ is formulated using the thermalized electron temperature $T_e$ and density $n_e$ which are used to calculate the corresponding Debye length, defined as \cite{chen1984introduction}
\begin{equation}
    \lambda_D = \sqrt{\frac{\epsilon_0 k_BT_e}{n_ee^2}},
\end{equation}
with $e$ being the elementary charge and $k_B$ the Boltzmann constant. 
The validity of utilizing the screened Coulomb interaction with the Debye length ($\lambda_D$ ) depends on whether it is much smaller than the size of the system $L$ and if there are enough electrons $N$ in the interval of the Debye length to make the screening statistically significant. Given these two conditions, the screened Coulomb interaction can be used for all length scales only when $\lambda_D << L$ and $N >> 1$ \cite{chen1984introduction}. For our simulations, the first condition is always fulfilled, but not necessarily the latter. For these cases, Stewart and Pyatt \cite{stewart1966lowering} developed theory for calculating the lowering of ionization potentials given a system of charged ions and free electrons which can be used for all values of electron temperatures and densities.
Alternatively, the nonlinear Poisson equation can be solved self-consistently, where the electron gas adapts to the ion charge density \cite{PhysRevE.70.051904} and provides means for an inhomogeneous electron density. This is important for systems containing heavy atoms, as they can attract more electron density compared to lighter atoms, thus leading to a more inhomogeneous density of free electrons. 

In the strongly coupled regime with high density/low temperature, the ion sphere (IS) model is more physical. The basis of the ion sphere model is that each ion is surrounded by enough bonded and free electrons to keep the neutrality of the ion sphere. Given an atom with net-charge $Q$ and a free electron density $n_e$, the radius of the sphere is calculated as
\begin{equation}
    R = \left(\frac{3Q}{4\pi n_e}\right)^{1/3}.
\end{equation}
The potential outside the radius $R$ from the ion is zero, while inside different models for the screening potential can be used, like the uniform electron gas model \cite{li2019analytical, MURILLO19981}
\begin{equation}
    \phi_{IS}(r) = \frac{Q}{r} - \frac{Q}{2R}\left(3-\frac{r^2}{R^2}\right).
\end{equation}
The ion charge, electron density and temperature of the plasma dictate whether it is more suitable to use the Debye screening or ion-sphere model. These depend on the pulse parameters and the sample. In some cases, the system is in a state between the two extremes. This motivates the use of a hybrid model that incorporates the ion-sphere model for some radius $r < r'$ (strongly coupled regime) and the Debye screening model for $r> r'$ (weakly coupled regime). By matching the ion sphere and the screened Coulomb  model at the boundary $r=r'$ and both their first and second derivatives, one can derive the following form of the electrostatic potential \cite{MURILLO19981}
\begin{equation}
%\begin{align}
      \phi (r) = \phi_{IS}(r) + \phi_{D}(r)=  \\ \frac{c_0}{r} + c_1 - c_2r^2 +  \frac{c_3}{r}\exp(-\frac{r}{\lambda_D}).
%\end{align}
\label{hydridpotential}
\end{equation}
%\frac{ze^2}{2R^3}
The coefficients $c_0, c_1, c_2$ and $c_3$ are given by the boundary conditions and are dependent on the free electron density, temperature and ion charge. The boundary point $r'$ where $\phi_{IS}(r')= \phi_{D}(r')$ is defined as \begin{equation}
    r' = \lambda_D (((\frac{R}{\lambda_D})^3+1)^{1/3}-1),
    \label{transition-radius}
\end{equation} 
%$r' = \frac{1}{\kappa} (((\kappa R)^3+1)^{1/3}-1)$
and is dependent the Debye screening length $\lambda_D$. 

By utilizing the form of the electrostatic potential presented in equation (\ref{hydridpotential}), the interaction can dynamically adapt to the current state of the plasma, encompassing parameters such as ion charge, electron temperature and density. This feature renders the use of this model suitable for studying any system that undergoes a transition into a plasma. %By using the form of the electrostatic potential in equation (\ref{hydridpotential}), the interaction can adapt to the current state of the plasma (ion charge, electron temperature and density), which makes using this model suitable for any system studied that transitions into a plasma. 
As all interactions between the atoms are formulated by electromagnetic force, screening due to free electrons will affect all interactions between two atoms. We therefore also modulated the Lennard-Jones (LJ) interaction according to the Debye
length, similar to the Debye model, 
\begin{equation}
    \phi_{LJ}(r_{ij}) = \bigg (\frac{c_{12}}{r_{ij}^{12}} - \frac{c_6}{r_{ij}^6} \bigg ) \exp \bigg (\frac{-r_{ij}}{\lambda_D}\bigg ).
\label{Screend Lennard Jones}
\end{equation}
\begin{figure}[!h]
     \centering
     \begin{subfigure}[b]{0.235\textwidth}
         \centering
         \includegraphics[width=\textwidth]{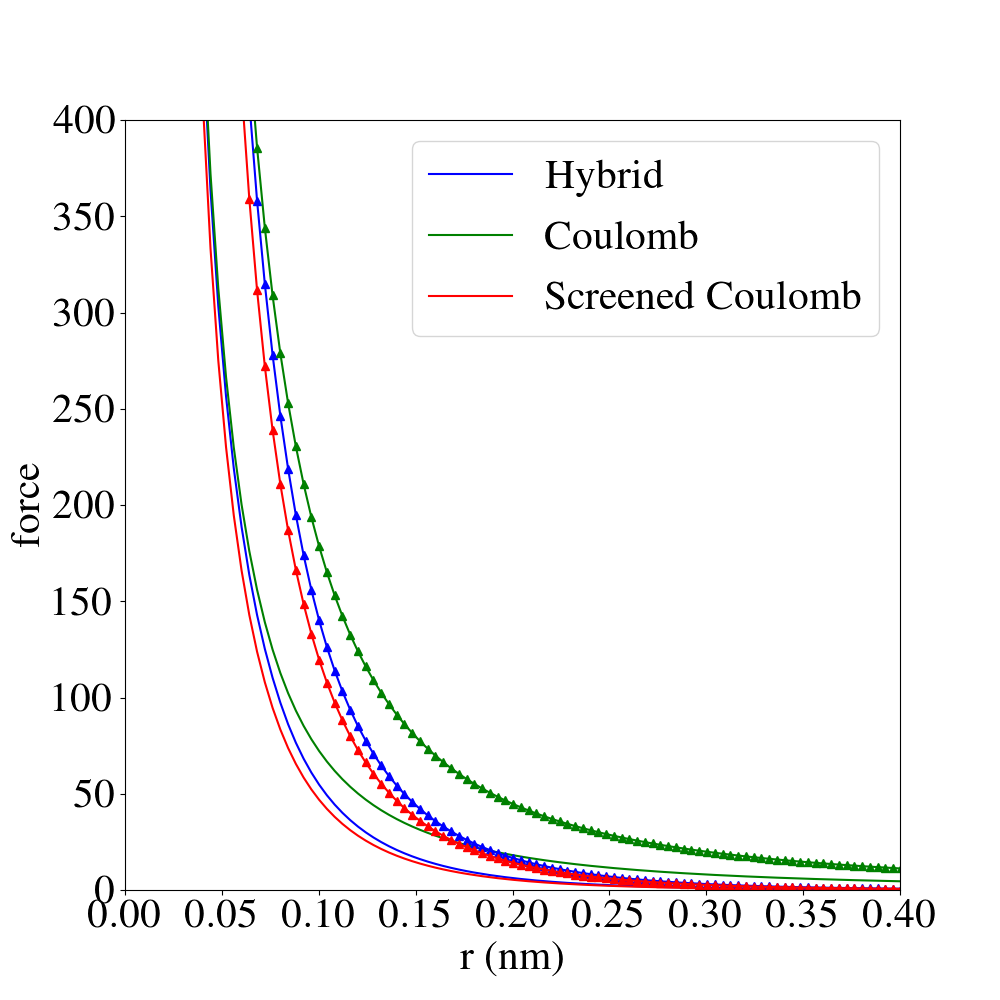}
        \caption{}
         \label{fig:1e18_hybrid_screening}
     \end{subfigure}
     %\hfill
     \begin{subfigure}[b]{0.235\textwidth}
         \centering
         \includegraphics[width=\textwidth]{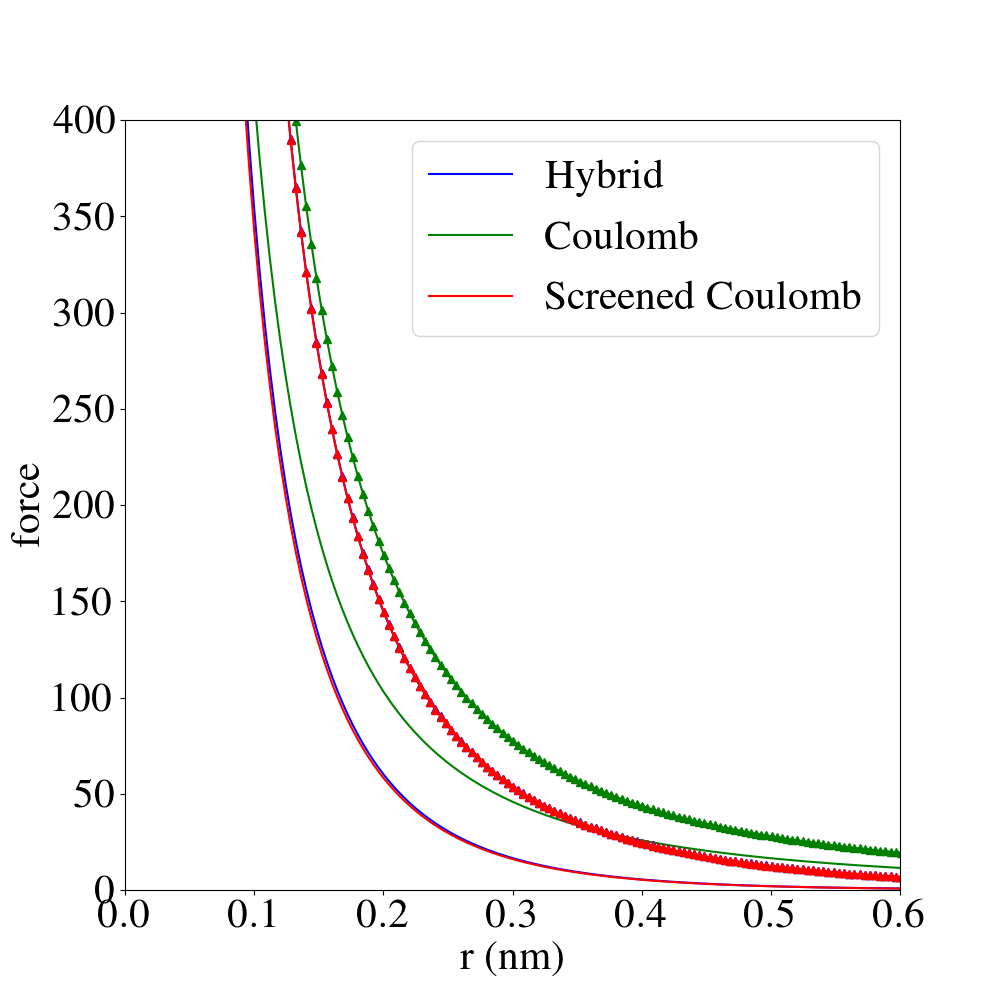}
        \caption{}
       \label{fig:1e19_hybrid_screening}
     \end{subfigure}
     \caption{The electrostatic forces given by the hybrid screening potential in equation (\ref{hydridpotential}), compared to the Debye Coulomb screening model in equation (\ref{SCreend coulomb potential}) and the normal Coulomb force. Data extracted from a simulation of bulk water exposed to a 75 fs flat XFEL pulse with 6860 eV photons and intensity a) $10^{18}$ W/cm$^2$ and (b) $5 \times 10^{19}$ W/cm$^2$.  Two time-points during the simulation are shown, 10 and 75 fs. $r'$ corresponds to the transition point between the ion-sphere and Debye screening model calculated from equation (\ref{transition-radius}). The solid lines correspond to the forces after 10 fs, and the solid-dotted lines are the same forces calculated at 75 fs.}
     \label{fig:hybridpotential}
\end{figure}
%In figure (\ref{fig:hybridpotential}), the hybrid screening potential, the Debye screening potential and the classical Coulomb interaction are compared. 

In figure (\ref{fig:hybridpotential}) we compare the hybrid screening potential, Debye screening potential, and classical Coulomb interactions to illustrate the impact of how reduced the screened potentials are compared to the classical Coulomb interaction. %This is done to asses how reduced the screened potentials are compared to the classical Coulomb interaction.
Furthermore, we are interested in understanding when the hybrid screening and Debye screening models become equal. We note that for a lower intensity of $10^{18}$ W/cm$^2$ in figure (\ref{fig:1e18_hybrid_screening}) that there is a difference between the hybrid force and the Debye screened model. Both are heavily reduced compared to the case of no screening, however the Debye model generally screens more compared to the hybrid model. For the higher intensity in figure (\ref{fig:1e19_hybrid_screening}), the difference is reduced. Additionally, we note that the transition point $r = r'$ is shifted to higher values favouring the IS model with time for the lower intensity (going from $r'=0.0816$ nm  to $r' = 0.0865$ nm), while it does the opposite for higher intensities (going from $r'=0.0602$ nm  to $r' = 0.0235$ nm), thus favoring the DS model. 

\subsection{Force field modelling}
The force field used in the MD simulations will differ depending on the sample studied. For instance, for metal clusters, non-bonded and bonded interactions  such as the Morse potential and a potential for angles suffice to describe the dynamics. This is due to the rather isotropic structure of the system which allows for the use of only spherically symmetric two-body potentials between atoms. For biological macromolecules like proteins, the structure is more complicated, as they contain, in addition to Coulomb and Lennard-Jones, interactions describing bonds, angles, dihedrals and cmaps. 
\begin{comment}
    In these force fields, atoms which are interacting through bonded terms, usually have their non-boned interactions excluded. For instance, if atom $i$ has neighbours $i+1$ and $i+2$ their interactions requires a quantum mechanical treatment due to overlapping molecular orbitals, which means they do not interact trough the Lennard-Jones term. When we simulate proteins, we have seen god success with excluding the Lennard-Jones term for the neighbours $i+1$ and $i+2$. The bond between $i$ and $i+1$ is described by a Morse potential and the potential between $i$ and $i+2$ is formulated by an angular term centered on $i+1$. 
\end{comment}
%Since this term is not dependent on distance in regular MD simulations, we turn it off when $r_{i,i+1}$ or $r_{i+1,i+2}$ is larger than 1.5 the bond length.
The bonded potentials in the force field are valid for the native structure, and they will be changed dynamically during the simulations due to the change in electronic occupation of the ionized atoms. Based on the mean charge $\Bar{z}$ with the total of $N$ atoms involved, we scale the bonded interaction with a value $c$ between 0 ($\Bar{z}=1$) and 1 ($\Bar{z}=0$), calculated by \begin{equation}
    c = 1-\frac{\Bar{z}}{N}.
\end{equation}
This is supported by previous DFT simulations of charged amino-acids \cite{aminosyror}, which show that the Coulomb interaction will dominate over bonded interactions as charge increases. The motivation for using these potentials even though the sample is defined as a plasma is that every atom has a certain probability of being neutral for the entire simulation period. In this case, the bonded terms for these atoms should still remain. 
For atoms which are bonded in the native structure, we use Morse potentials in order to allow for bond breaking. The potential is defined as,
\begin{equation}
 V_{\text{Morse}}(r_{ij}) = D_e \exp(-(r_{ij}-b_e)^2),
    \label{Morse}
\end{equation}
where the parameters $D_e$ (dissociation energy), $b_e$ (equilibrium distance) and the current distance $r_{ij}$ depend on the two atoms $i$ and $j$ involved. For these atom pairs, the LJ interaction is omitted, but we still apply the screened Coulomb interaction. 
Partial charges present in the native force field are not used, since charges are localized to a specific atom
when bond breaking occurs. %Keeping track of how they are distributed between atoms after ionization is complicated and would require ab-initio simulations to accurately incorporate.  

We omit molecular bonds between atoms which are not initially bonded in the native system. Two charged atoms in a plasma are unlikely to form bonds since the potential energy surface becomes dissociative (i.e. there is no minimum energy) for just a few removed electrons, which occurs early in the simulation. Thus, the potential energy surface of two arbitrary charged atoms approaching each other from infinity is best described by a non-bonding potential, such as the LJ interaction. For this potential, the equilibrium distance is usually larger than molecular bond lengths, since they
are based on the effective ionic radius. This radius will be reduced as electrons are removed from the atom, due to the reduction of the extension of the electron cloud. We have therefore computed new LJ parameters for each combination of charge states between atoms, based on the values for the neutral parameters, which are changed dynamically in the simulation. The new parameters were calculated by performing density functional theory calculations of all possible charge states of the atoms using the RSPt [16] package and extracting the electron density, which was compared to the undamaged electron density. Alternatively, we have used a scaling factor $c$ for reducing the equilibrium distance, based on the net-charge of the atoms. Given atom $i$, with net-charge $q_i$ and atomic number, $Z$ the factor $c \in [0, 1]$ is determined as
\begin{equation}
    c_i = 1-\frac{q_i}{Z}.
\end{equation} 
Standard classical MD codes rely on using cut-offs for the non-bonded interactions to reduce the computational time. Since the phase of the matter we probe is pushed into a plasma, the cut-off that we use can be greatly reduced due to the screening from the free electrons. We can compute a suitable non-bonded cut-off by determining the maximum charge states observed in the CR simulations, and search for the distance which provides a force smaller than some tolerance. This in combination with parallelization through domain decomposition, has allowed us to explore large samples on longer time-scales.

With all bonded and nonbonded potentials involving interactions between atoms that are defined in the total potential $V(\textbf{r}) = V_{\text{bonded}}(\textbf{r}) + V_{\text{nonbonded}}(\textbf{r}) $, we can determine the corresponding forces. 
For propagating the dynamics of the system, we use the leap-frog algorithm in  \textsc{Gromacs}  to solve Newton's equation
\begin{equation}
    m\frac{d^2\textbf{r}_i}{dt^2} = - \nabla V(\textbf{r}),
    \label{integrator_md}
\end{equation}
where $\textbf{r}_i$ is the position of atom $i$. 
Equation (\ref{integrator_md}) includes the effect of the free electrons in the screening of the Coulomb interaction between the ions. However, the free electron distribution can also interact with the ions directly, which could affect their trajectories. This is valid for a weakly coupled system where the Coulomb interaction is heavily reduced compared to the kinetic energies. By utilizing the Langevin equation (stochastic dynamics), we introduce two additional terms in the right-hand side of equation (\ref{integrator_md}) as 
\begin{equation}
    m\frac{d^2\textbf{r}_i}{d t^2} = - \nabla V(\textbf{r})- m_i\gamma  \frac{d\textbf{r}_i}{d t} + \textbf{f}_i(t), 
    \label{langevin}
\end{equation}
where the first term is a friction force and the second a stochastic Gaussian force\cite{apol2010gromacs, molina2022molecular, cimarron} with the property  
\begin{equation}
    \bigl \langle f(t)f(t+t') \bigr \rangle = 2m_i\gamma_ik_BT_e\delta(t'),
\end{equation}
%$\bigl \langle f(t)f(t+t') \bigr \rangle = 2m_i\gamma_ik_BT_e\delta(t')\delta_{ij}$
which means that calculation of the Gaussian force at each time-point for atom $i$ is independent of any other time-point. The constant $\gamma$ with unit $s^{-1}$ is calculated using the electron-ion coupling $g_{e-ion}$ from the CR calculation and the number density $\rho_{N} $ as $\gamma = \rho_N g_{e-ion}$. $T_e$ is the electron temperature retrieved from the CR calculations. 
%All quantities that that the coupling are dependent on change with time in the CR calculations, but an an average during the computation time is used as input to the MD simulation. 
The model for the electron-ion coupling parameter $g_{e-ion}$ will dictate the energy transfer between the ions. In our model, we use an average value for the coupling parameters as it does not change much during the simulation for each atomic species. The electron temperature is dynamically changed at every time-step and it is read from the CR calculations.

In our model, we are limited by the size of the single particle as the CR code assumes no escaping electrons and we simulate them in vacuum. %We therefore need to find a size where a low amount of electrons escape. 
We estimated the minimum size of a single particle based on the pulse parameters used, to trap all produced free electrons. For the same net charge density, a larger radius would result in a reduced number of escaping electrons. For a radius $R$ and positive charge density $\rho$, the potential for a charged sphere at distance $r$ from its center is \cite{hau2004dynamics}
\begin{equation}
    \phi(r) = \frac{\rho}{2\epsilon_0}\bigg(R^2-\frac{r^2}{3}\bigg).
    \label{eq:critical_potential}
\end{equation}
A free electron will escape if its kinetic energy is larger than the potential from the sphere at the boundary $r=R$. Thus, we get the following condition:
\begin{equation}
    E_{kin} > \frac{\rho R^2}{2r\epsilon_0}.
    \label{threshold_energy}
\end{equation}
The first 1-3 fs in the CR  simulation, the charge density $\rho$ will not be large enough to trap any free electrons. As the net charge will rapidly increase and reach the critical potential in this time-scale, it allows for the approximation that the low number of escaping electrons does not affect the results.

\subsection{Coherent scattering calculations }
We determined the displacement of the system from the MD simulations weighted by the distribution of the probe pulse and the charge state distribution of the atoms. Given a Gaussian pulse shape \cite{Jonsson:xh5043} $g(t)$ with duration $T$, satisfying $\int_0^T g(t)dt = T $, and the time-evolution of the fraction of remaining bound electron in an atom (with atomic number Z) $f(t)=1-\frac{\Bar{z}}{Z}$, we determine the weight $w(t)$ as 
\begin{equation}
    w(t) = \int_t^{t+dt}\bigg(\frac{g(t)f(t)}{\int_0^T g(t)f(t)dt} \bigg)dt,
\end{equation}
where $dt$ is the time-step of the MD simulation and $T$ the total pulse duration. The real space weighting is an approximation of the average probed structure, motivated by the fact that the phases are not well-defined. %The displacement plot provides means of determining suitable pulse lengths for conducing SFX experiments on proteins.

In order to compare with experiment, an observable needs to be computed. We propose that the fragments of the sample could be studied, with for instance mass spectrometry or reaction microscopes available at SQS at the European XFEL. Alternatively, the signal from coherent scattering in diffraction experiments can be studied. For bulk water, we computed coherent scattering using the structure factor $S(q)$ defined as \cite{soper2007joint}
\begin{equation}
    S_{\alpha \beta}(q,t) =  1 + \frac{4\pi \rho_{0}}{q}  \int_{0}^{\infty} r(g_{\alpha \beta}(r,t) - 1 ) \sin (qr) \:  \mathrm{d}r \,
    \label{S_ab},
\end{equation}
where $g_{\alpha, \beta}$ is the radial distribution function (RDF) between atomic species $\alpha$ and $\beta$ available from the MD simulations. The scattered intensity is then computed numerically as 
\begin{equation}
    I= \int_0 ^T g(t) f(q, t)^2 S(q, t) dt
    \label{eq:intensity}
 \end{equation}
where $f(q, t) = \int \rho (r) e^{i\textbf{q}\cdot\textbf{r}}d\textbf{r}$ is the form-factor defined as the Fourier transform of the atomic electron density at time $t$ and $g(t)$ is the normalized pulse profile. This atomic electron density is computed using density functional theory calculations in RSPt, for arbitrary electronic configurations. The form-factor $f_{\alpha}$ for atomic species $\alpha$ at  time $t$ is weighted by the all the different electronic configurations $j$ with weight $w_j$ and form-factor $f_j$ as
\begin{equation}
    f_{\alpha} (q, t) = \frac{\sum_j w_{\alpha, j}(t)f_{\alpha, j}(t)}{\sum_j w_{\alpha, j}(t)},
    \label{weighted_formfactor}
\end{equation}
where the weight are extracted from the CR calculations.

\subsection{Samples studied}
In the results section we present studies of a single particle (cluster of atoms), a liquid (bulk water) and a protein crystal (lysozyme). We began by studying non-thermal heating of water as given by an XFEL. This was motivated by our previous experiments on non-thermal heating and scattering of water using the XFEL at the Linac Coherent Light Source (LCLS) \cite{beyerlein2018ultrafast}. The work also included simulations which were compared to the experiments. There we used an older model which did not directly couple the CR calculations to the MD. In this work, we used the same PDB structure for the water and CR calculations, but now instead used our updated hybrid CR/MD model to simulate non-thermal heating and scattering. 

Our model was additionally compared to an SFX X-ray pump X-ray probe experiment on lysozyme crystals\cite{nass2020structural}. There, the disulfide bond length as a function of probe delay was studied to follow the effects of ionizing radiation on the structure. We used a solvated lysozyme structure and simulated the evolution of the disulfide bonds in the Cystine amino-acids due to the different X-ray intensities and probe delays, comparable to the ones used in the experiment. The time-delays of the probe used was 0, 20, 40, 60, 80 and 100 fs.

\begin{figure}[h]
     \centering
     \begin{subfigure}[b]{0.4\textwidth}
         \centering
         \includegraphics[width=\textwidth]{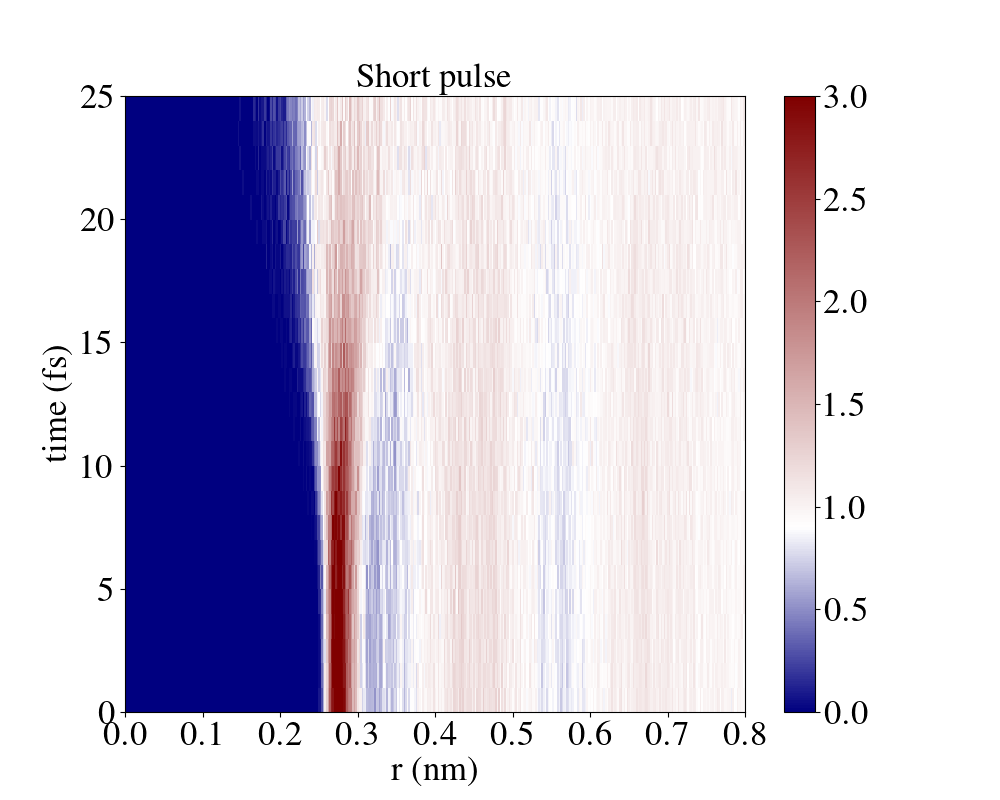}
        \caption{}
         \label{fig:short_rdf}
     \end{subfigure}
     %\hfill
     \begin{subfigure}[b]{0.4\textwidth}
         \centering
         \includegraphics[width=\textwidth]{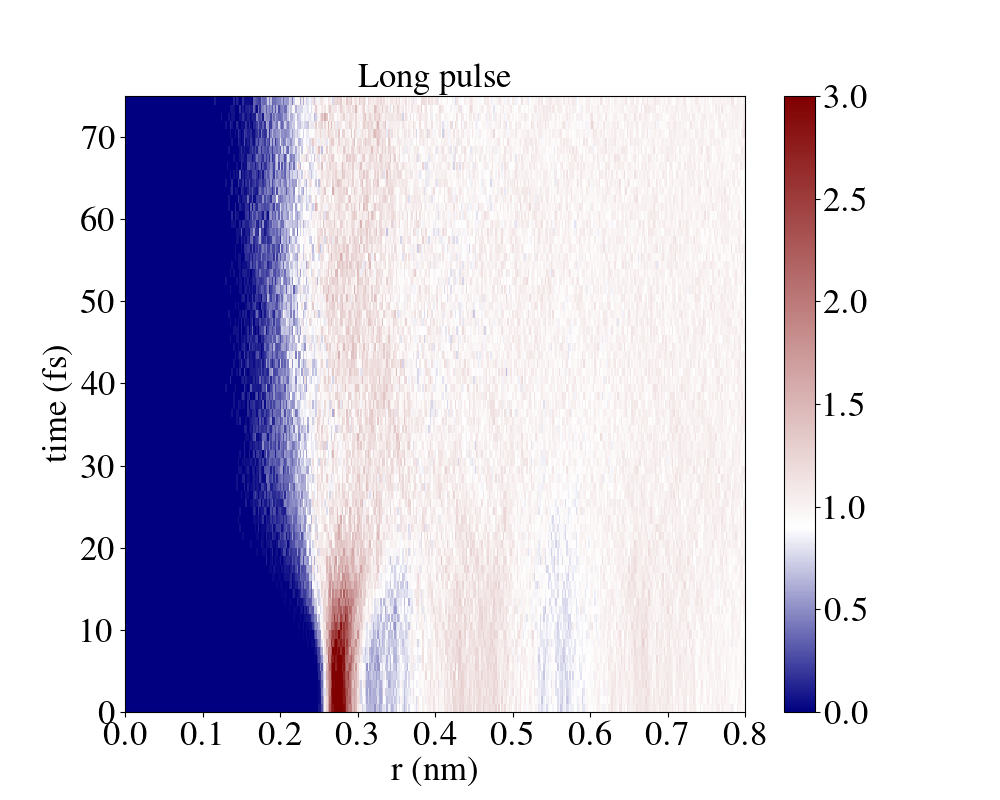}
        \caption{}
             \label{fig:long_rdf}
     \end{subfigure}
     \caption{RDF of bulk water for pulse duration of a) 25 fs (short) and b) 75 fs (long). In XFEL simulations, we alter the LJ parameters based on the reduction of bound electrons, the ionic radius is reduced and the atoms can therefore reach closer compared to the neutral case. }
      \label{fig:RDF-water}
\end{figure}

For the single particle case, we simulated a large methane cluster consisting of approximately 45000 atoms. This was studied in the context of single particle imaging with XFELs. The size of the system was chosen based on the relationship between the number of escaped electrons and the radius of the droplet. The potentials used for this system are the screened Coulomb and LJ interaction described by equations (\ref{SCreend coulomb potential}) and (\ref{Screend Lennard Jones}), where the parameters for the LJ interaction are taken from the TIP3P water model in the CHARMM\cite{MacKerell1998a} force field. For the bonded interactions, we used Morse potentials and a harmonic angle term.  
We prepared the methane cluster assuming a liquid density of $0.4$ g/cm$^3$. Given the target density and the molar-mass of $M=16$ g/mol we could compute the number of methane molecules for any volume.  \textsc{Gromacs}  was then used to, in a randomized way, insert the corresponding number of molecules in the given volume, based on the atoms' van der Waal radii. Before production simulations, we minimized the energy using steepest descent.

\section{Results: model validation and comparison to experimental data}

\subsection{Non-thermal heating of bulk water}
\begin{figure}[h]
    \centering
    \includegraphics[width=0.45\textwidth]{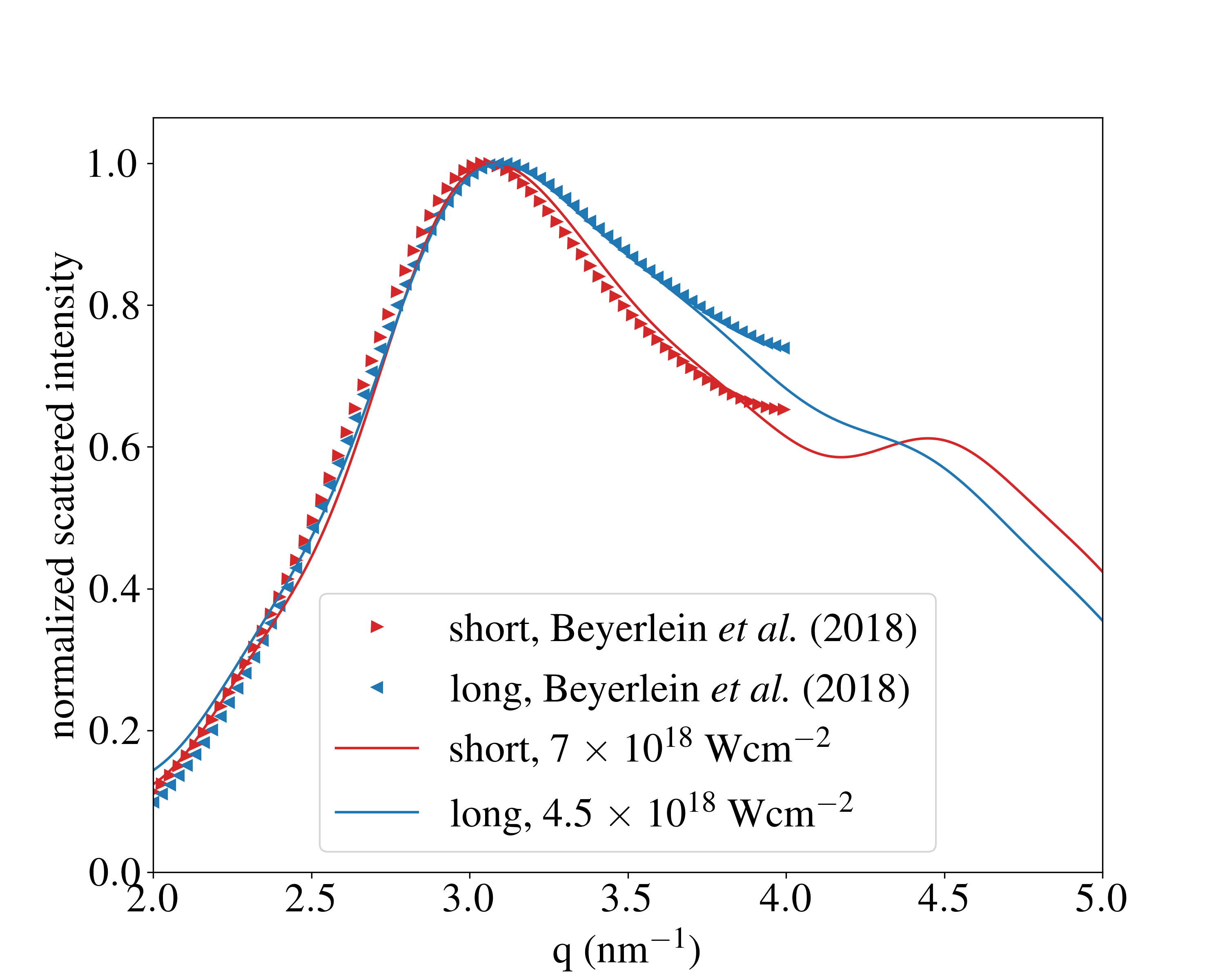}
\caption{Scattering of bulk water using the newly developed model compared to experimental data \cite{beyerlein2018ultrafast}. The red curve is the short 25 fs pulse, with intensity $7 \times 10^{18}$ Wcm$^{-2}$ and the blue is 75 fs long with an intensity of $4.5 \times 10^{18}$ Wcm$^{-2}$. }
    \label{fig:nonthermal_water_scattering}
\end{figure}

We compute the scattered intensity for bulk water as it is probed by an intense XFEL pulse. The simulations are compared to previous experiments and simulations of non-thermal heating of water 
\cite{beyerlein2018ultrafast}. The real space dynamics of the atoms is shown in the radial distribution function in figure (\ref{fig:RDF-water}). It can be seen that a short pulse of 25 fs in (\ref{fig:RDF-water}a) alters the structure of water, by increasing the disorder of the system. This is manifested through the new length scales being occupied in the RDF. This disorder starts to be significant at about 15 fs, which results in that during a large part of the pulse duration, the short pulse coherently scatters from a near native sample. This provides both the first and second peak in the scattered intensity as seen in figure (\ref{fig:nonthermal_water_scattering}), encoding the frequency of the relative distance between O-O coordination peaks. For the long 75 fs pulse in (\ref{fig:RDF-water}b), the disorder is introduced around 20 fs, and the pulse therefore largely scatters from a sample that is far from the native structure. This results in that the second peak at $q \approx 4.5$ nm${^{-1}}$ in figure (\ref{fig:nonthermal_water_scattering}) is lost.

\begin{figure}[h]
     \centering
     \begin{subfigure}[b]{0.235\textwidth}
         \centering
         \includegraphics[width=\textwidth]{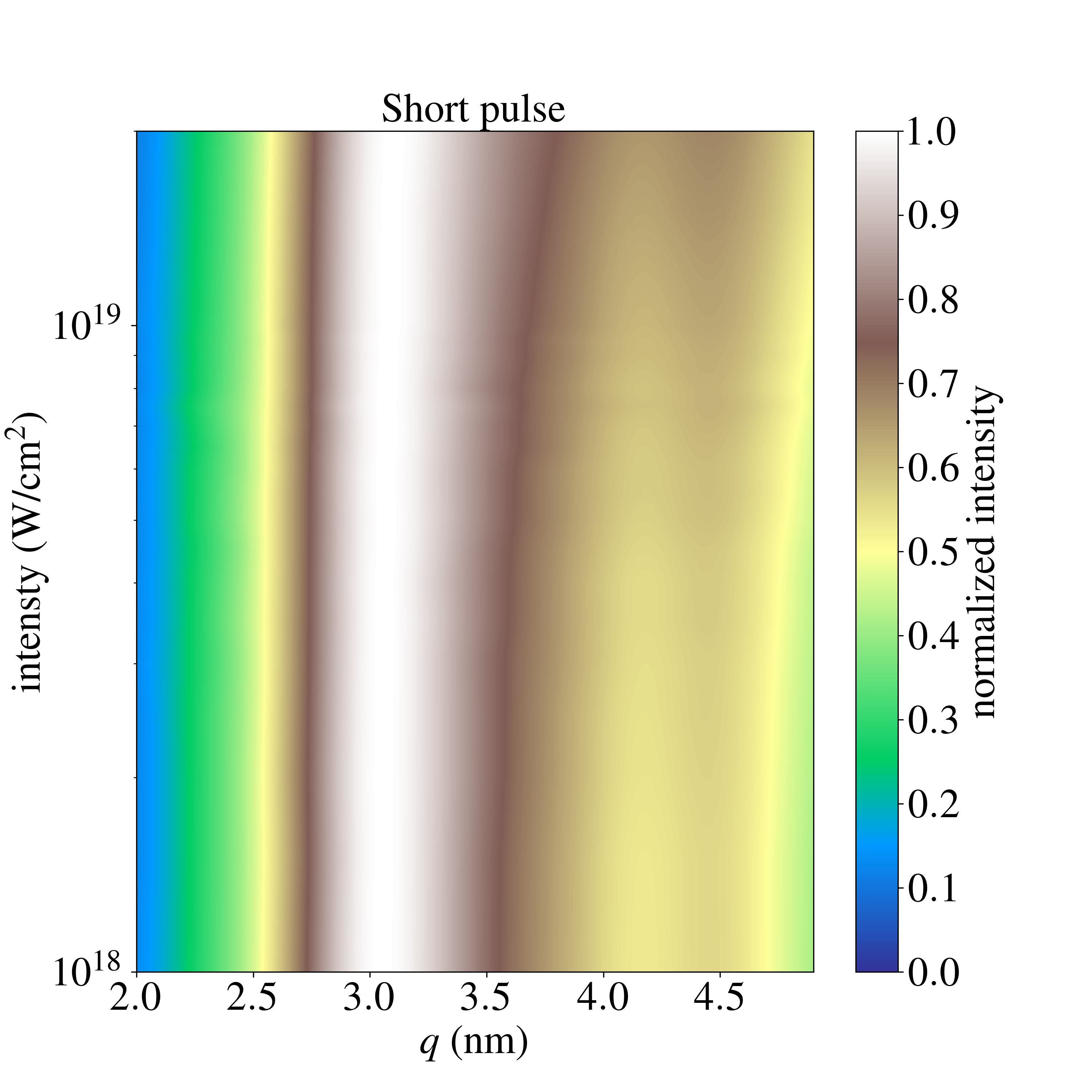}
        \caption{}
         \label{fig:short_pulse_integrated}
     \end{subfigure}
     %\hfill
     \begin{subfigure}[b]{0.235\textwidth}
         \centering
         \includegraphics[width=\textwidth]{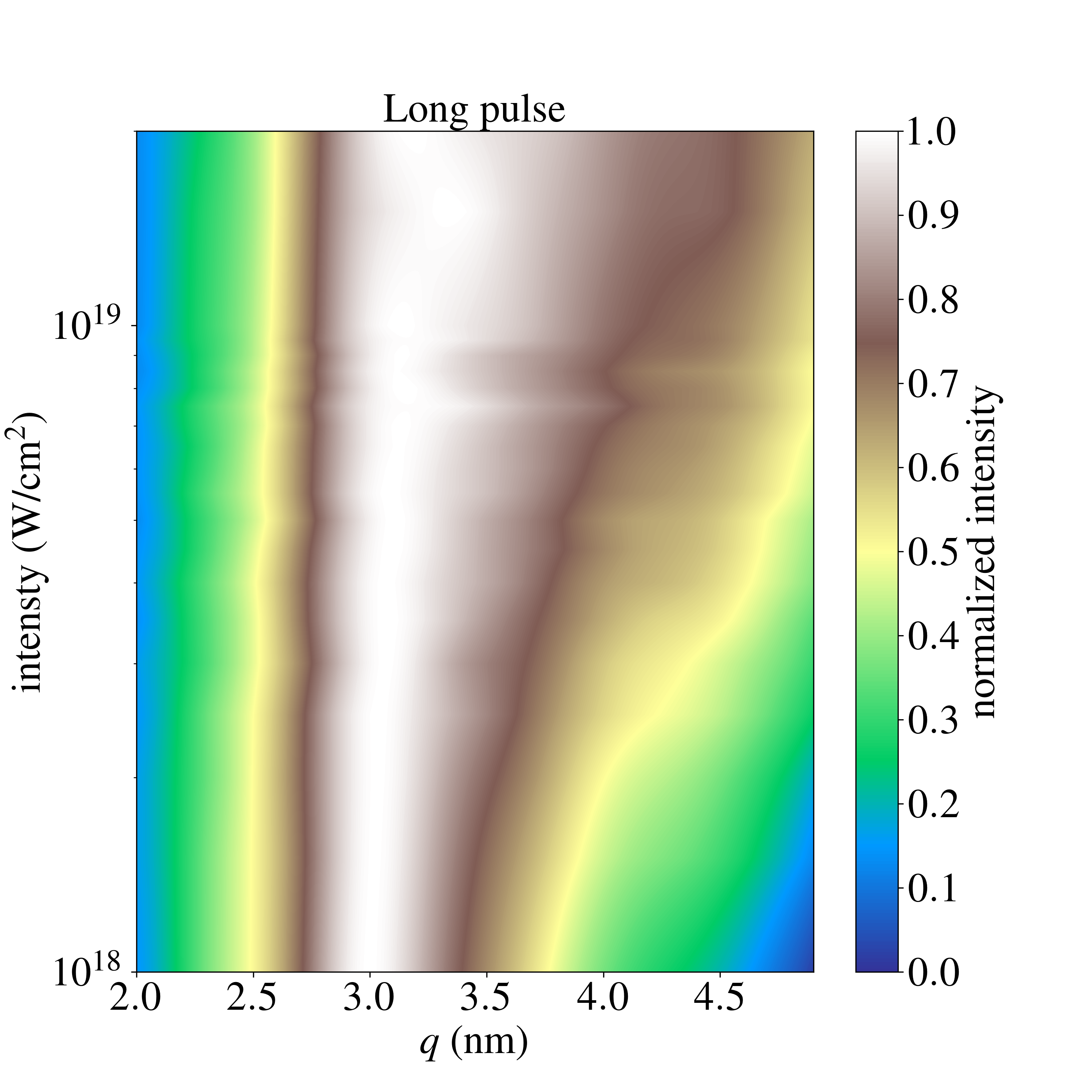}
        \caption{}
         \label{fig:long_pulse_integrated}
     \end{subfigure}
     \caption{Time-integrated scattering as a function of momentum transfer and the incident XFEL intensity for the a) short 25 fs pulse and b) long 75 fs pulse. }
      \label{all_integrated_intensities}
\end{figure}

Using equation  (\ref{eq:intensity}) we calculate the scattered intensity from bulk water as its probed by the XFEL, shown in figure (\ref{fig:nonthermal_water_scattering}).
%We note that the relationship between the short and long pulse is the same for the theoretical calculation as in the experiment. 
To reproduce the behaviour of the experimental curves using our current model, we need to reduce the fluence compared to the values reported in the paper that is compared, and reduce the ratio between the intensities.
%, while keepingthe ratio (3x) between the short and long pulse. 
The intensity in XFEL experiments can vary substantially between shot-to-shot such that the fluence in the oridinal study is an average one. Additionally, both the spatial and temporal distribution of the intensity can vary within a shot, making it difficult to accurately determine the fluence that the sample is exposed to. 
We simulated a range of different XFEL intensities, to understand how increasing the incident intensity leads to changes in the structure and scattering signal. It can be seen in figure (\ref{all_integrated_intensities}) that as one increases the XFEL intensity, the momentum transfer values above approximately $3.5$ nm$^{-1}$ start to accumulate scattering intensity. This means that the structure is changing, where new length scales are occupied by the oxygen atoms. One can note that the longer pulse sees a stronger change compared to the short pulse. 

%\newpage

\subsection{Disulfide bond breaking in femtosecond crystallography with X-ray pump X-ray probe}
\begin{figure}[h]
    \centering
    \includegraphics[width=0.45\textwidth]{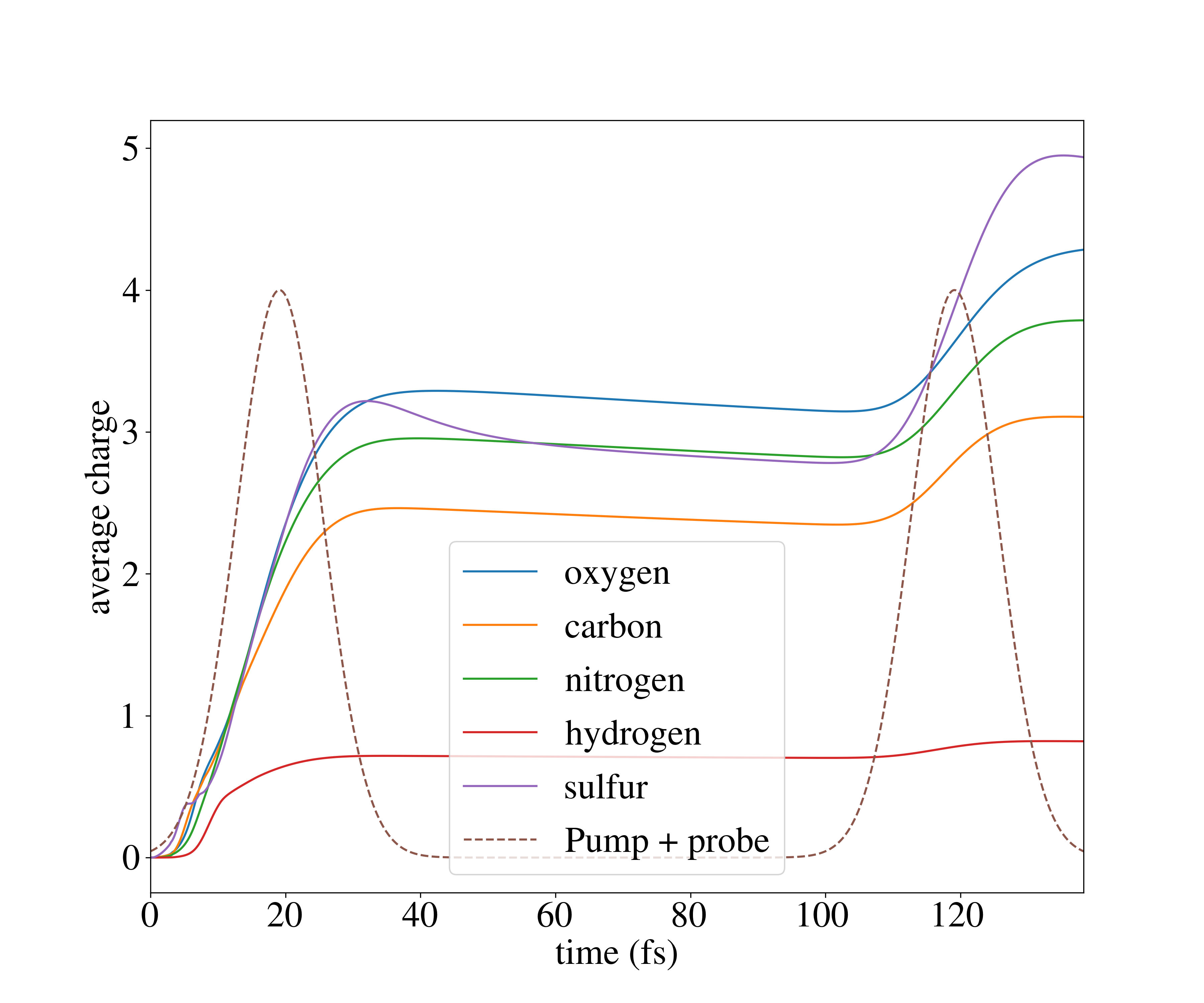}
\caption{Example of the average charge of the atoms as a function of time for the lysozyme protein crystal, induced by a 15 fs FWHM X-ray pump and X-ray probe scheme. An intensity of $6.75 \times 10^{18}$ Wcm$^{-2}$ with a pump photon energy of 7152 eV and a probe energy of 7072 eV. }
    \label{fig:average_charge_hipip}
\end{figure}

We compare our model to a recent experiment which imaged the disulfide bond breaking in a lysozyme protein crystal given by an X-ray pump X-ray probe set-up \cite{nass2020structural}. Similar to the experiment, we model both the pump and probe as a Gaussian with a FWHM of 15 fs. We simulated several different intensities, in the ranges reported in the experiment. An example of the average charge for all atoms calculated for a 100 fs probe delay is in figure (\ref{fig:average_charge_hipip}).

\begin{figure}[!h]
\centering
\includegraphics[width=0.45\textwidth]{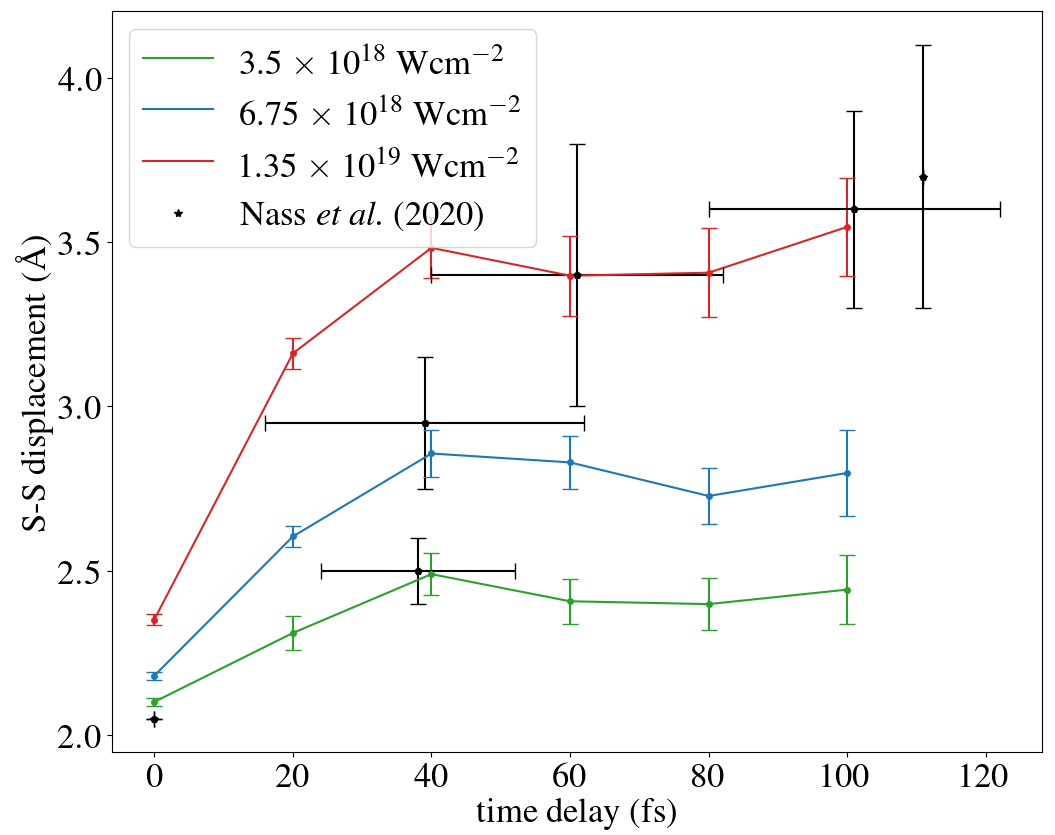}
\caption{Pulse weighted relative displacement between the disulfide bond (S-S) as a function of probe delay, computed for multiple intensities. The data is averaged over all disulfide bonds in the Cystine amino-acids. Nass \textit{et al.} (2020) represents experimental data\cite{nass2020structural}, as presented in the paper.}
\label{fig:pump_probe}
\end{figure}

The pulse-averaged disulfide (S-S) bond distance as a function of intensity and probe delay is shown in 
figure (\ref{fig:pump_probe}), together with the experimental data points \cite{nass2020structural}.
It can be noted that the simulations show a good agreement with the experiment as a displacement between the disulfide bond reaches saturation, but depending on the time-delay a different intensity might need to be used. According to our model, as the probe delay is increased, the intensity of the pulse should be higher or volume integration for the spatial pulse profile is needed \cite{xcalib}. 
%We theorize that the reason for this could be due to either that the experiment had larger intensities with higher time-delay. Alternatively, gating due to Bragg termination\cite{barty2012self, Caleman:15} could be a reason. 
We theorize the experimental intensities increased with longer time delays. An alternative explanation could be intensity gating from Bragg termination \cite{barty2012self, Caleman:15}.For long time-delays, the effect of the pump will have a longer time to manifest in the structure. Parts of the crystal that were exposed to the peak of the spatial profile of the pulse could be more heavily damaged. This would result in a disorder, and consequently the peak intensity would not contribute to the Bragg spots. Instead, a diffuse scattering background would be added to the scattering data. The part of the crystal that is exposed to the less intense part of the spatial profile would instead have their crystalline coherence maintained through reproducible movement in the unit-cell. This would then contribute to the Bragg spots, and later provide the reconstruction of the system. A result which could support this, is that we note that for higher probe delays the variance of the reconstructed disulfide distances increases in the experiment. We have seen in our simulations, that as the probe delays and intensities increase, so does the variance of the determined distance.    

We note that the largest change in displacement is for the 20 fs probe delay. For larger delays, the displacement plateaus. This is due to the ion caging affect, where neighbouring atoms provide a force which limits the expansion. This caging affect has been observed for disulfide bonds in X-ray pump, X-ray probe experiments \cite{nass2020structural}. Our results show that larger probe delays provide the approximately the same mean displacement as shorter, but with a larger variance. 

For short probe delays, we see that there should be some correlated motion, which is understood from the relatively small variance compared to the longer delays, which are very varied. Thus, no coherence would be seen for the longer pulses due to the large variance. In shorter probe delays, the variance will put a limit on the achievable resolution. Using our model, we can therefore provide estimations of the maximum attainable resolution by computing the variance.

\section{Results: application of the model and predictions}
\subsection{Fragmentation dynamics of a methane cluster}

\begin{figure}[h]
    \centering
    \includegraphics[width=0.45\textwidth]{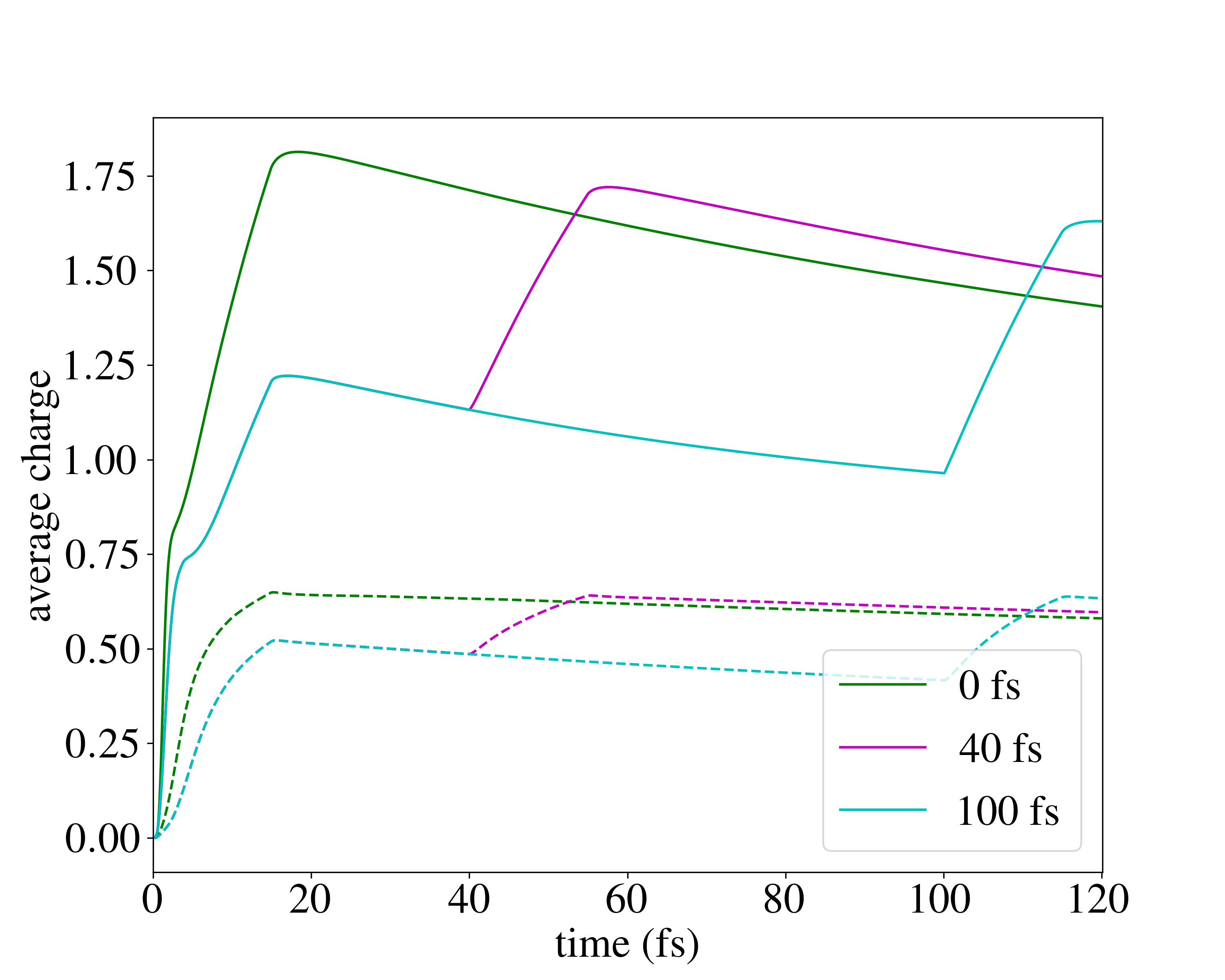}
\caption{Average charge of carbon (filled lines) and hydrogen (dashed lines) atoms as a function of time for a methane cluster at liquid density 0.4 g/cm$^{3}$ induced by an X-ray pump X-ray probe setup. The pulse has an intensity of $10^{19}$ W/cm$^2$, a FWHM of 15 fs, photon energy of 8 keV and time-delays of 0, 40 and 100 fs. }
    \label{plasma-data-methane}
\end{figure}

In order to study radiation damage in SPI, we envision that an X-ray pump X-ray probe scheme could be used. The pump would initiate the radiation damage in the sample, and a delayed probe would then image the dynamics. We use relevant beam parameters for SPI, where the X-ray laser has an intensity of $10^{19}$ Wcm$^{-2}$, photon energy of 8 keV, a 15 fs duration and a flat top shape time and no volume integration. The sample used is a methane cluster of 45000 atoms, serving as a model system for biological macromolecules such as proteins and viruses. Previous experiments have been done on fragmentation dynamics in methane clusters \cite{PhysRevA.86.033201} and recently there has been an increased interest in coherent imaging of nano-clusters, which turn into a plasma during the imaging \cite{10.1063/4.0000006}. Figure (\ref{plasma-data-methane}) shows the average charge  of the system due to the X-ray pump X-ray probe calculations.

\begin{figure}[!h]
    \centering
    \includegraphics[width=0.5\textwidth]{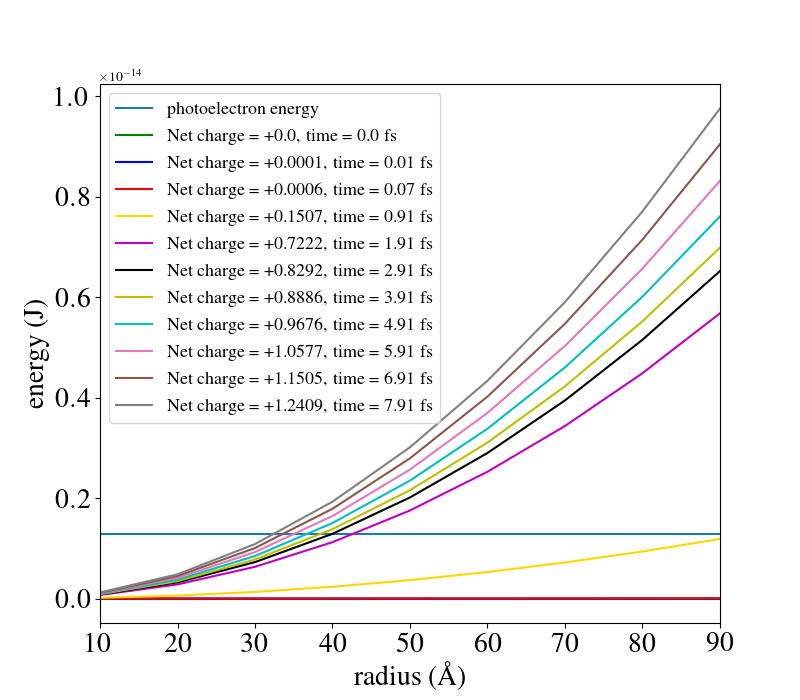}
\caption{Photoelectron energy compared to   electrostatic build-up from a methane cluster as the pulse interacts with the sample. The straight line corresponds to the photoelectron energy. The other lines are the electrostatic energies based on the average charges during a particular time-step, calculated using the right-hand side of equation (\ref{threshold_energy}).   }
    \label{fig:electrostatic_energy}
\end{figure}

For an atomic cluster exposed to an XFEL, the surface and the core of the system will behave differently. The surface of the system will be both less affected by the collisional processes and have a higher probability of moving into vacuum due to the reduced number of atoms in the near vicinity. For a small cluster, where the ratio of number of core atoms to surface atoms is small, assuming that a screened Coulomb interaction dictates the interaction would not be a good approximation. This is because atoms are less screened at the surface. For a large system, where the ratio of number of core atoms to surface atoms is large, the dynamics will be dominated by the screened potentials. 
The size of the cluster used in our simulations is motivated by the threshold energy of the cluster needed to trap the photo-electron produced. This is calculated based on the radius of the cluster and the positively charged density determined from the CR calculations, as described in equation (\ref{threshold_energy}). Given figure (\ref{fig:electrostatic_energy}), by using a cluster of size 5.5 nm, the system would build up an electrostatic energy large enough to trap the photo-electron within 2 fs. This makes it appropriate to use the CR simulations, which assumes no escaping electrons, the entire duration of the simulation.

\begin{figure}[h]
    \centering
    \includegraphics[trim={5cm 1cm 0 0},clip,width=0.5\textwidth]{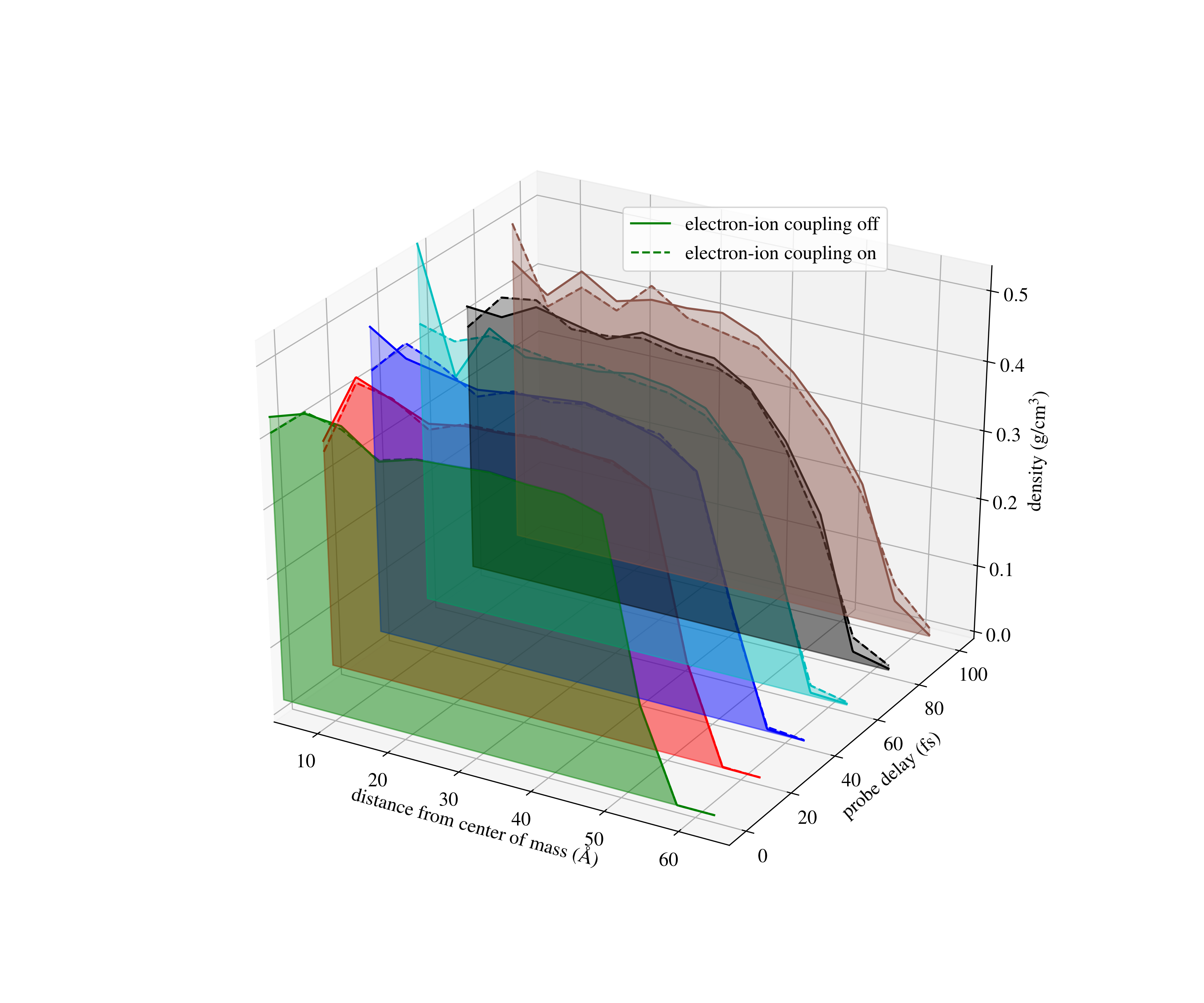}
\caption{Mass number density including hydrogen and carbon of a 5.5 nm methane cluster as a function of distance, probe delay and distance from the center of mass. The electron density for each probe delay is averaged during the probe pulse duration. A resolution of 5 \AA{} was used to sample the radius of the sphere.  The dotted line corresponds to a simulation with electron-ion coupling, and solid line is without electron-ion coupling. }
    \label{fig:density_evolution_methane}
\end{figure}

Our investigation is designed according to an experimental procedure to study X-ray induced damage dynamics. This is done using an X-ray pump X-ray probe scheme, where the probe is delayed by 0, 20, 40, 60, 80 and 100 fs. The probe pulse averaged electron density as a function of distance from its center of mass (COM) is then recorded and shown in figure (\ref{fig:density_evolution_methane}). In the figure, we compare the atomic number density as a function of distance to the center of mass, with and without electron-ion coupling.
We note that as time evolves, more outer layers of the cluster are removed, indicated by the reduction in the density close to the surface and occupation of larger distances from COM than the initial radius of the cluster. The density  is largely intact up to the 100 fs probe delay. 
It can be concluded that the effect of the electron-ion coupling using a resolution of 5 \AA{} provides a negligible difference.

\begin{figure}[!h]
     \centering
     \begin{subfigure}[b]{0.475\textwidth}
         \centering
         \includegraphics[width=0.9\textwidth]{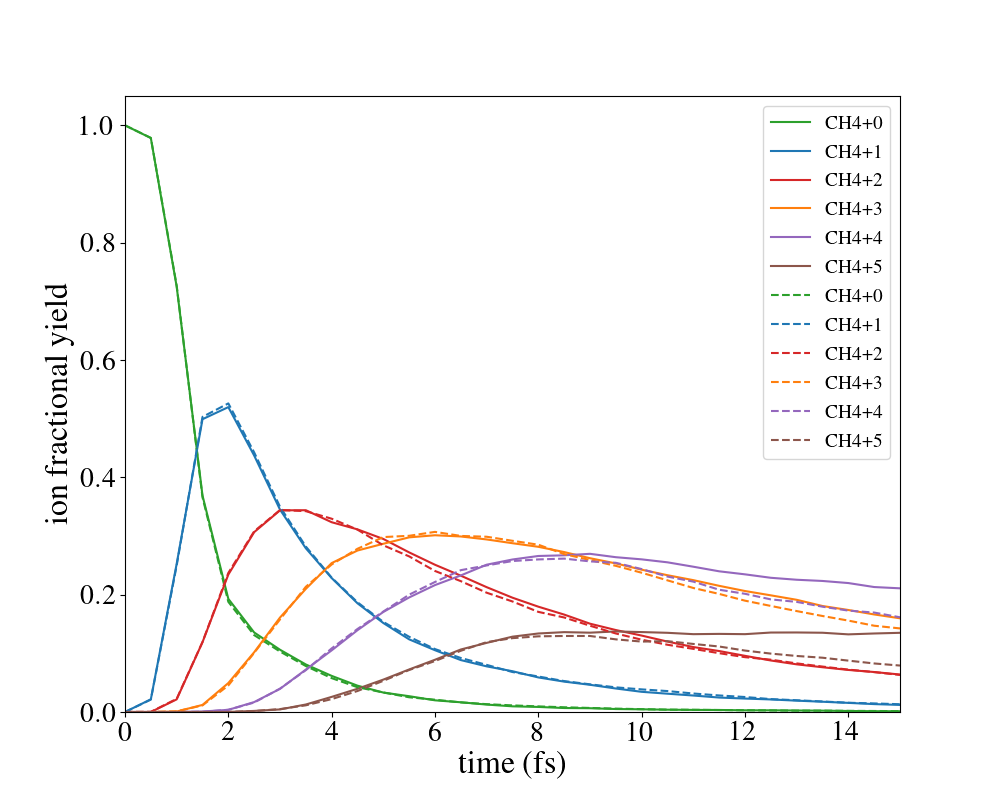}
        \caption{}
         %\caption{Average charge of carbon (filled lines) and hydrogen (dashed lines) atoms as a function of time for a methane cluster at liquid density 0.4 g/cm$^{3}$ induced by a 15 fs XFEL pulse at 92 eV. }
         \label{fig:kinetic_energy_unscreened}
     \end{subfigure}
     %\hfill
     \begin{subfigure}[b]{0.475\textwidth}
         \centering
         \includegraphics[width=0.9\textwidth]{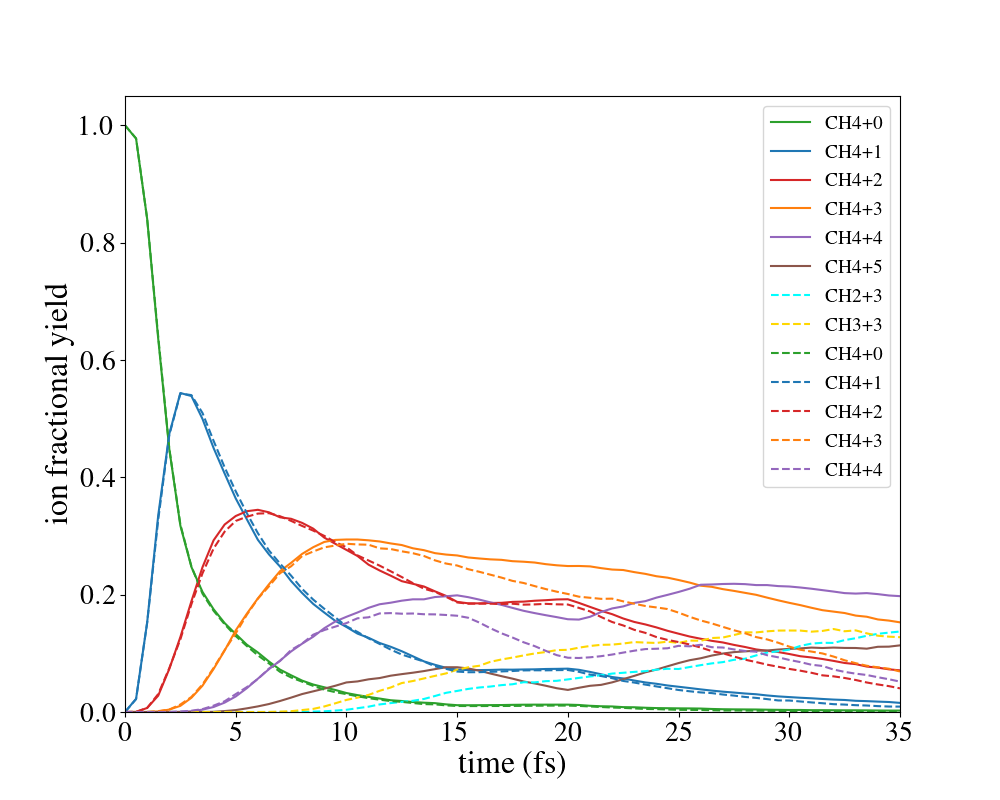}
        \caption{}
         %\caption{Debye length used for screening the Coulomb interaction as a function of time for a methane cluster at liquid density 0.4 g/cm$^{3}$ induced by a 15 fs X-ray pulse at 92 eV.}         \label{fig:kinetic_energy_screened}
     \end{subfigure}
      \begin{subfigure}[b]{0.475\textwidth}
         \centering
         \includegraphics[width=0.9\textwidth]{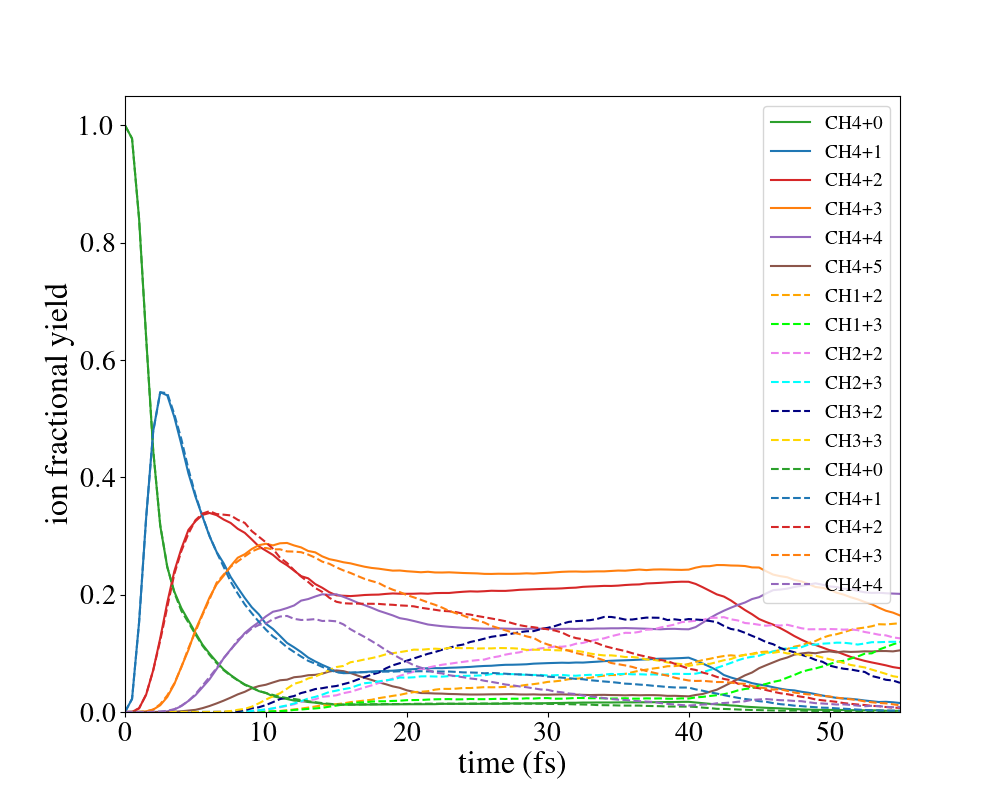}
        \caption{}
         %\caption{Debye length used for screening the Coulomb interaction as a function of time for a methane cluster at liquid density 0.4 g/cm$^{3}$ induced by a 15 fs X-ray pulse at 92 eV.}         \label{fig:kinetic_energy_screened}
     \end{subfigure}
     \caption{Difference in resulting  fractional ion yield in a methane cluster due to a 15 fs XFEL pump pulse with XFEL probe delays 0 fs in a), 20 fs in b) an 40 fs in c). Solid lines correspond simulations with electron-ion coupling, and the dotted lines are without electron-ion coupling.  }
      \label{fig:ion_yield}
\end{figure}

We further investigated the different fragments that are produced during the coherent imaging of the system. Fractional ion yields are show in figure (\ref{fig:ion_yield}), which provide information regarding which fragments with the specific net-charge that exist. It can be seen that the two models are similar for the early probe delays of 0 and 20 fs, but they deviate  for longer time-scales. Even though the ion yield are significantly different comparing the shortest time-delay and the longest of 100 fs, the resulting atomic number density that probe sees will be relatively similar, at least at the computed resolution of 5 \AA.

\subsection{The effect of electron-ion coupling on SPI and SFX}
Structural information can be retrieved from the fragmentation dynamics when a sample is exposed to radiation. The kinetic energies of the different particles, their trajectories and the mass to charge ratio are possible ways of analyzing the ions. The properties of the sample will dictate how these values behave, which provides means of retrieving structural information. Efficient simulation tools to explore long time-scales are important to understand fragments created during x-ray imaging.

%\section{The effect of electron-ion coupling on dynamics}
The dynamics of the ions is dependent on the transfer of energy from the pulse to the ions. In the MD simulation, their kinetic and potential energy can change due to the increase in the charge states of the atoms, which increases the Coulomb interaction. Another pathway for energy transfer is from the free electrons, which should be accounted for in the MD simulation. The ion-electron coupling rate affects how fast the energy exchange between the particle occurs, which could have an impact on the dynamics of the ions. For simulations that should run for a long time for comparison with fragmentation experiments, it is important to asses the effects of the X-ray laser on short ($\leq 100$ fs) time-scales and the heating from the electrons on long pico-second time-scales. 

To determine the effect of electron-ion coupling, we compared the resulting scattered intensity for bulk water (data not shown). We note that the difference between the data is negligible. Thus, for the  pulse lengths of 25 and 75 fs, there is not enough time for the oxygen atoms to move to detect a difference between the two implementations. However, for longer time-scales, much larger than the pulse length, the electron-ion coupling can provide a difference in the resulting dynamics. To asses the relevance of electron-ion coupling, we compute the fractional ion yield for methane clusters as a function of time and probe delay for a simulation with and without electron-ion coupling, shown in figure (\ref{fig:ion_yield}).  

For higher intensities, the time-point where the distribution of ion yield starts to differ occurs earlier. The change in ion yield for the simulation without electron-ion coupling saturates shortly after pulse, where the rate of change is small. With coupling however, the distribution continuously changes. This is because after the pulse has terminated, the simulation without coupling will not be pumped with energy any more, while in the other case electrons will continue to transfer their kinetic energy to ions. This enables the creation of new mass/charge fragments through the ions having higher kinetic energies, since the charge states change very slowly after the pulse. 

\begin{figure}[h]
     \centering
     \begin{subfigure}[b]{0.475\textwidth}
         \centering
         \includegraphics[width=\textwidth]{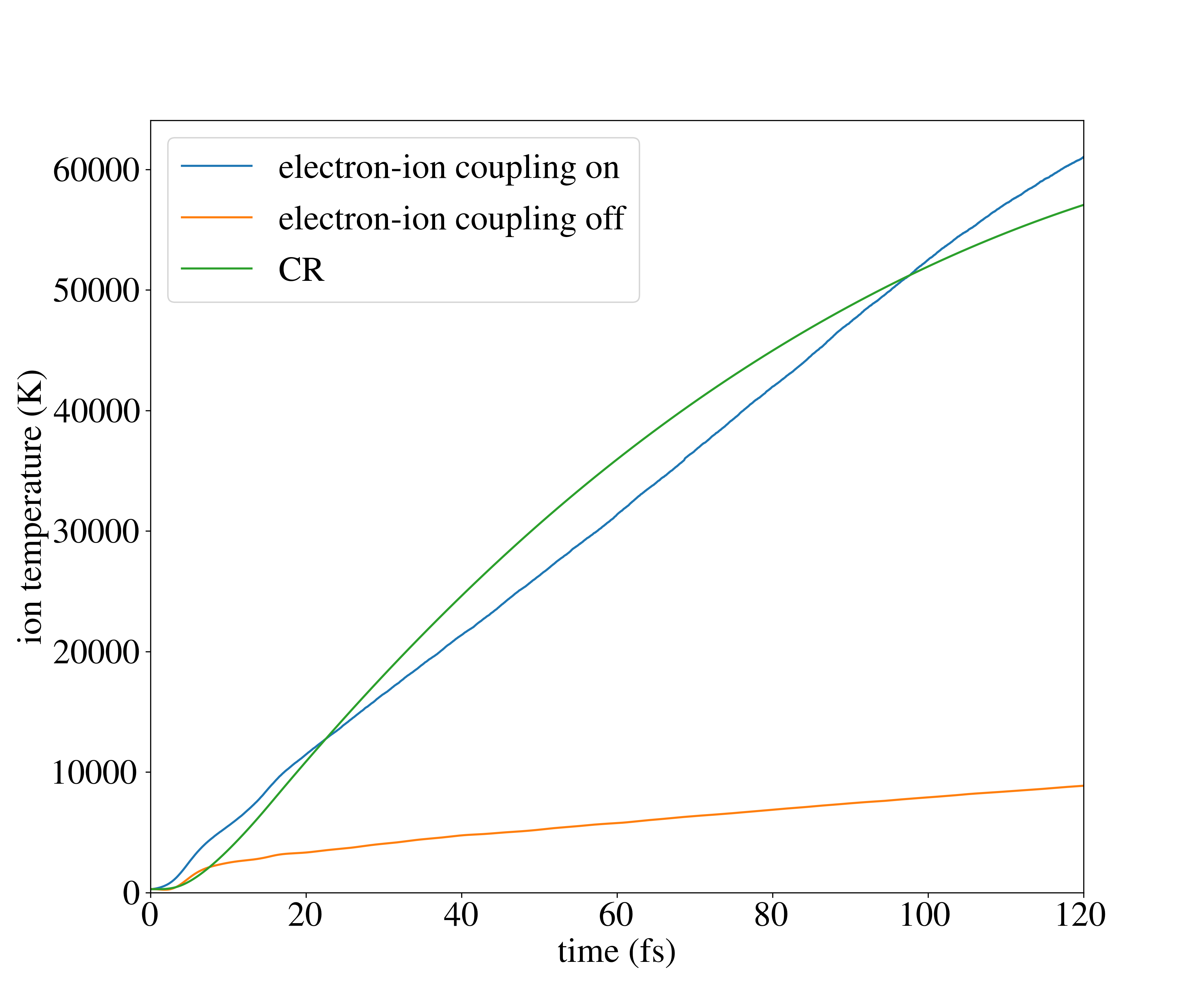}
        \caption{}
         \label{fig:0_fs_probe_delay}
     \end{subfigure}
     %\hfill
     \begin{subfigure}[b]{0.475\textwidth}
         \centering
         \includegraphics[width=\textwidth]{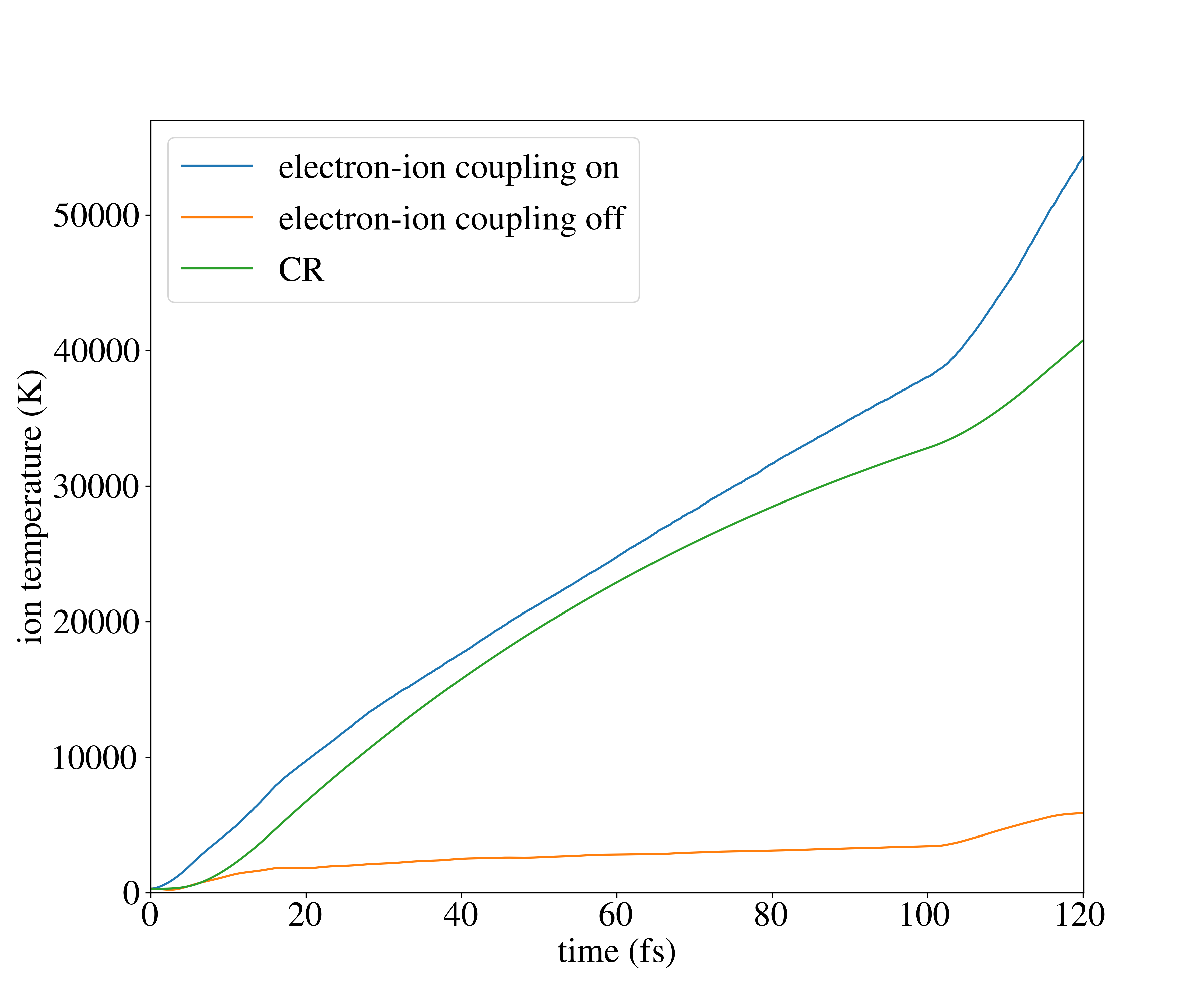}
        \caption{}
             \label{fig:ion_temperature}
     \end{subfigure}
     \caption{Ion temperature for the methane cluster given by the CR simulation in  \textsc{Cretin}  for 0 fs probe delay in a) and 100 fs delay in b).   }
      \label{fig:methane_ion_temperature}
\end{figure}

The conclusion is that for the 15 fs pulse, which would be detected in an SPI experiment on a photon detector through coherent diffraction, is not affected incorporating electron-ion coupling or not. The pulse has the largest effect on the dynamics through the change of charge states and the free electron distribution. 
Even though electron-ion coupling does not affect on time-scales relevant for coherent imaging with X-ray laser, the thermodynamics of the system in the MD simulation will not reproduce the CR ones if the coupling is omitted. We calculated the ion temperature as given by the CR calculations in  \textsc{Cretin}, and MD using with and without electron-ion coupling in figure (\ref{fig:methane_ion_temperature}). The MD contains hybrid screening as developed in the theory section and charge states from the CR data. It can be concluded that including electron-ion coupling provides much better agreement with the CR simulations. It is clear that for long time-scales, on the order of picoseconds, electron-ion coupling should be included for a spherically expanding plasma.

\section{Conclusion and outlook}
We have presented a hybrid method to compute photon-matter interaction for X-ray lasers, implemented in the classical MD code  \textsc{Gromacs}  with an addition of collisional-radiative calculations that take into account dynamical changes in the ion and electron dynamics. \textsc{MolDStruct} is able to simulate a great variety of samples in different 
conditions that are explored using high-intensity X-ray lasers. 

We have compared our theoretical method to two experiments, on bulk water and protein crystals, and we see a quantitatively good agreement. Therefore, we believe that the code  will be important in order to asses the limits of experiments and to understand if experiments with sensitive samples have been affected by radiation damage. 
It is particularly important to asses structural changes due to damage in time-resolved experiments, since these changes might be interpreted as biological function. The code is presently being used to compare theoretical results with several SFX experiments.

We envision that the model could be directly compared to an experiment where a small molecule crystal \cite{schriber2022chemical} is studied using an X-ray pump X-ray probe scheme at an XFEL. By having a sample containing heavier atoms which scatter strongly, the atomic movement can be followed as a function of probe delay, by analyzing the scattering data using Patterson functions \cite{rupp2009biomolecular}. The Patterson function is a real space quantity which is possible to extract from both experiment and the model presented here.  

The strength of the model is that we can access long time scales in large-scale samples. We are able to include processes such as electron-ion coupling,  dynamically modify forces with screening and changes in the ionization. The model allows the study of local damage, which is important for protein radiation-damage sensitive parts, like in metalloproteins. One limitation of the model is that it treats the electrons as a gas that instantly thermalize and does not follow the explicit dynamics of the electrons. The fast electron dynamics can be modelled with Monte Carlo methods, while the thermal electrons could still be modelled as a gas\cite{xtant3}. Another limitation is that \textsc{MolDStruct} is currently suitable for large samples which can be described by a plasma phase, where the electrons are trapped and lead to a non-thermal heating of the sample. For simulating small systems such as small proteins, we are developing a hybrid Monte Carlo/Molecular Dynamics model, which is applicable for modelling amino-acids and proteins.

\begin{acknowledgments}
We thank Erik Marklund, Filipe Maia, Olof Jönsson and the Biophysics network at Uppsala university for fruitful discussions.
%Talis Uelisson da Silva is acknowledged for providing topology files for running the high-potential iron-sulfur protein molecular dynamics simulations in GROMACS. 
All calculations were done on the Davinci computer cluster provided by the
Laboratory of Molecular Biophysics, Uppsala University.
Project grants from the Swedish Research Council (2018-00740, 2019-03935) are acknowledged, the Röntgen-Ångström Cluster provided by the Swedish Research Council (2021-05988) and the Helmholtz Association through the Center for Free-Electron Laser Science at DESY. 
\end{acknowledgments}

\newpage

%\nocite{*}
\bibliography{aipsamp}% Produces the bibliography via BibTeX.
%\bibliography{references}% Produces the bibliography via BibTeX.

\end{document}